\pdfoutput=0
\documentclass[iop]{emulateapj}
\usepackage{graphicx}
\usepackage{amsmath}
\usepackage{amssymb}
\usepackage{array}
\usepackage{multirow}
\usepackage{rotating}
\usepackage{txfonts}
\usepackage{epstopdf}

\input epsf

\def\lsim{\mathrel{\raise.3ex\hbox{$<$\kern-.75em\lower1ex\hbox{$\sim$}}}}
\def\gsim{\mathrel{\raise.3ex\hbox{$>$\kern-.75em\lower1ex\hbox{$\sim$}}}}

\def\Gyr{\,{\rm Gyr}}
\def\Myr{\,{\rm Myr}}
\def\yr{\,{\rm yr}}
\def\yri{\,{\rm yr^{-1}}}

\def\pc{\,{\rm pc}}
\def\mic{\,\mu{\rm m}} 
 
\def\Kpc{\,{\rm kpc}}
\def\Mpc{\,{\rm Mpc}}

\def\Msol{M_{\odot}}

\def\erg{{\,\rm erg}}
\def\cmm2{{\,\rm cm^{-2}}}
\def\cm2{{\,{\rm cm}^2}}
\def\cmm3{{\,{\rm cm}^{-3}}}
\def\gcmm3{{\,{\rm g\,cm^{-3}}}}
\def\kms{\,{\rm km\,s^{-1}}}

\newcommand{\SFR}{{\rm SFR}}

\newcommand{\GMLR}{{\rm GMLR}}
\newcommand{\SFH}{{\rm SFH}}
\newcommand{\ml}{{\textit{ml} }}
\newcommand{\nml}{{\textit{nml} }}
\def\tgc{{\tau_{\rm gc}}}

\begin{document}

\title{Fuel Efficient Galaxies: Sustaining Star Formation with Stellar Mass Loss}

\author{Samuel N. Leitner and Andrey V. Kravtsov}
\affil{Department of Astronomy \& Astrophysics, The
  University of Chicago, Chicago, IL 60637 USA} 
\affil{Kavli Institute for Cosmological Physics and Enrico
  Fermi Institute, The University of Chicago, Chicago, IL 60637 USA} 

\smallskip
\begin{abstract}
We examine the importance of secular stellar mass loss for fueling
ongoing star formation in disk galaxies during the late stages of
their evolution. For a galaxy of a given stellar mass, we calculate
the total mass loss rate of its entire stellar population using star
formation histories derived from the observed evolution of the
$M_{\ast}$-star formation rate relation, along with the predictions of
standard stellar evolution models for stellar mass loss for a variety
of initial stellar mass functions. Our model shows that recycled gas from
stellar mass loss can provide most or all of the fuel required to
sustain the current level of star formation in late type galaxies.
Stellar mass loss can therefore remove the tension between the low gas
infall rates that are derived from observations and the relatively
rapid star formation occurring in disk galaxies. For galaxies where
cold gas infall rates have been estimated, we demonstrate explicitly
that stellar mass loss can account for most of the deficit between
their star formation and infall rates.
\end{abstract}


\maketitle
\smallskip

\section{Introduction}

An understanding of how galaxies get their gas is key to a complete
picture of galaxy formation and evolution. At all redshifts, star
forming galaxies appear to be living fast and dangerously: they are
observed to be converting their gas to stars at a rapid rate that
would lead to exhaustion of their cold gas reservoirs in about two
billion years without a fresh supply of
gas \citep[][]{Kennicutt1998,Leroy2008,Bigiel2008,genzel_etal10,daddi_etal10,daddi_etal10b}.
Observations, however, also show that star forming galaxies are common
at all redshifts, and theoretical models indicate that star formation
rates of the most actively star forming systems are set by their cold
gas accretion rates \citep[e.g.,][]{dave_etal10,Bouche2010}. In
addition, since the average neutral hydrogen density in the universe
has dropped by only $50\%$ over 10 billion years in the face of much
higher gas consumption rates, the cold gas supply must be getting
continuously replenished \citep[e.g.,][]{bauermeister_etal10,
Prochaska2009,putman_etal09a}.

Despite its importance, the mechanism by which gas is replenished in
nearby galaxies is a major puzzle \citep[see, e.g.,][for
reviews]{Sancisi2008,putman_etal09a}. Clouds of neutral hydrogen are
observed around the Milky Way, M31 and a number of other nearby
galaxies \citep[e.g.,][]{blitz_etal99,Wakker2007,Miller2009,Thilker2004},
but both the fate of these clouds and the fraction of their gas that
they deposit in their host's disks are uncertain. Direct searches in
nearby groups of galaxies have failed to uncover massive populations
of clouds of neutral hydrogen \citep{pisano_etal07}. Nevertheless,
estimates for the accretion rate due to high velocity HI clouds (HVCs)
have been made for a handful of galaxies, and typical values range
from $10\%$ to $20\%$ of the current SFR of the parent
galaxy \citep[see][for a review, and table~\ref{table:gasbudget}
below]{Sancisi2008}.

The accretion of gas rich satellite galaxies provides a second
potential source of fresh gas for massive star forming host
galaxies \citep[e.g.,][]{Sancisi2008}. Larger satellites have the
potential to carry fresh gas to the star forming disk of their hosts
when they merge, and such a scenario is likely in the case of M31 and
M33 for example \citep{putman_etal09}. However, due to the gas
stripping that occurs during the most common minor mergers, satellites
will tend to deposit their gas at large radii, well beyond the radius
of their host's star forming
disk \citep{Peek2009a,Grcevich2009}. \citet{Sancisi2008} estimate that
some nearby galaxies accrete gas from satellites at a rate of $\sim
10-20\%$ of the SFR of their hosts, although this number could be
significantly lower on average \citep[][]{Kauffmann2010}.

Such low gas accretion rates could imply that galaxies are presently
in the stage of exhausting their gas reservoirs and ceasing star
formation activity. However, this would leave questions about the gas
supply during the past several billion years, because the evolution of
the disk galaxy population is inconsistent with a complete shutdown of
star formation
\citep[e.g.,][]{bauermeister_etal10, Bell2007}.

So far, predictions of theoretical models as to how gas is delivered
to galaxy disks are quite uncertain. While theoretical studies predict
broadly that the average accretion rate of baryonic mass onto dark
matter halos should be several times their SFRs today
\citep[e.g.,][eq~1]{Dekel2009}, the timescale and route by which the
baryonic mass reaches galaxy disks themselves is difficult to predict
\citep[e.g.,][]{FaucherGiguere2011}. Most of the baryons accreted at
low redshifts are thought to encounter previously accreted material
and suffer shocks, instabilities, and fragmentation
\citep[e.g.,][]{Keres2009}; these events may mix the accreting gas
with hot halo gas before it has a chance to reach the star forming
disk where it is needed. Although such hot gas may cool at the
disk-halo interface and contribute to star formation at a later time,
this gas may not be immediately available.

Given that a large gas reservoir may surround galaxy disks, an
 inviting solution to the gas deficit problem is to tap this hot gas
 through mixing with colder gas on hot phase/cold phase boundaries.
 This mixing would elevate the cooling rate of the hot coronal gas,
 and allow it to sink onto the galaxy, thereby producing net
 accretion. Such mixing could occur in the galactic
 fountain \citep[e.g.,][]{Marinacci2010,Fraternali2008} and,
 intriguingly, might explain the decrease in angular momentum that is
 observed in gas with increasing height above the Milky Way disk.
 Mixing has also been proposed to cool gas out of the halo at the
 interface between the hot halo and dwarf satellites
\citep{Bland-Hawthorn2009}. Unfortunately, the predictions of these models are 
rather uncertain because magnetic fields, conduction and poorly constrained gas density, 
and temperature, can all have leading-order effects on the final mass of gas that cools 
during mixing \citep[e.g.,][]{Marinacci2010,Heitsch2009}.
Even without physical uncertainties,
the resolution required to track the detailed cooling of the
halo gas may be beyond what is achievable by
current simulations (e.g., an ability to resolve the cooling layer may be necessary), so the
thermodynamics and fate of the accreting gas are both difficult to
model and are poorly constrained.

Before exploring unconstrained and difficult to model mechanisms for
accretion, all other channels should be fully understood. A
significant internal gas channel -- the stellar population itself --
has not been rigorously explored in this context. Stellar populations
can shed a significant fraction ($\sim 30-50\%$) of their mass over
the Hubble time through secular processes, particularly via shedding
of the stellar envelopes during the Asymptotic Giant Branch (AGB)
phase \citep[e.g.,][]{Vassiliadis1993,Hurley2000,Groenewegen2007}. In
comparison, the Milky Way requires only $\sim 2\%$ of its stellar mass
to be returned to ISM per billion years to balance its gas consumption
by star formation. Stellar mass loss should be an accessible fuel for
star formation in disk galaxies because winds are typically injected
directly into the relatively dense interstellar medium (ISM) with cool
temperatures $<3\times 10^3K$, low velocities $\sim 10\kms$ and
largely in molecular form \citep[e.g.][and references
therein]{Knapp1990,Marengo2009,libert_etal10}.

Mass return by stellar populations is noted in the
 seminal study of \citep{Roberts1963}, which was the first
to discuss the transience of galaxy populations as a result of gas
 consumption. \citet{Sandage1986} and \citet{Kennicutt1994} both made
 the important observation that recycled material could extend gas
 consumption timescales in the nearby disk galaxy population. More recently,
 \citet{Blitz1997} has pointed to observations demonstrating
 that the mass of gas lost in the Milky Way could be a significant
 source of fuel. Nevertheless, in recent literature, stellar mass loss
 is generally not considered as part of the gas supply budget outside of
 some chemical enrichment studies.

Many codes for simulating galaxy formation have only recently begun to
incorporate stellar mass
loss \citep[][]{Katz1996,kravtsov_gnedin05,stinson_etal06} and recent
studies indicate that it can be an important physical ingredient,
significantly influencing galaxy
morphologies \citep{Agertz2010,Martig2010} and providing substantial
fuel source for continuing star formation at
$z=0$ \citep[][]{Schaye2010}.

The goal of this study is to carefully assess the potential importance
of stellar mass loss in fueling star formation of late-type galaxies
at low redshifts ($z<1$), by making use of observational constraints
from star formation measurements. To this end, we estimate the total
loss rates averaged over entire galaxies for empirically motivated
star formation histories (SFHs) and show that stellar mass loss could
be an important source of fresh gas for star formation at late epochs.
In particular, we explicitly demonstrate that gas returned to the ISM
by stars can provide most of the gas required to maintain current
level of star formation in a number of nearby galaxies, for which the
observed accretion rate of halo gas clouds appears to be insufficient
to fully resupply the gas reservoir.

To estimate the global stellar mass loss rate for a galaxy we need
a mass loss model that describes the mass loss rate of a single age stellar population as a function of time and
a star formation history describing the age distribution of all stellar populations in a galaxy. 
Mass loss modeling and its uncertainties are discussed in
\S~\ref{sec:modelingmassloss}. We calculate star formation histories
in \S~\ref{sec:globalobservedsfh} based on empirical measurements of
the slope and evolution of the relation between stellar mass and the
star formation rate of star forming disk galaxies. The resulting
global stellar mass loss rates that our model predicts are presented
in \S~\ref{sec:massloss_history}, where we also derive mass loss rates
for several nearby galaxies that are then compared to the observed
difference between star formation and cold gas infall. Our results and
conclusions are discussed in \S~\ref{sec:conclusions}. Finally, the
Appendix explores the extent to which recycled material might
accumulate over time based on the results of cosmological galaxy
formation simulations and simple models that reproduce those
simulations.

\section{Modeling Mass Loss}\label{sec:modelingmassloss}
 
To first order, the mass recycled by a single-age stellar population (SSP)
comes from mass that is lost by stars that have evolved through
the luminous, wind-driving, SNe, red supergiant and/or asymptotic
giant branch stages, minus the total mass in compact stellar remnants.
 The fraction of mass lost by a SSP
population at a given time $t$ after its birth is therefore approximately the fraction of the population's
stellar mass that was initally possessed by stars with masses above
the main sequence turn off $m\geq m_{\rm to}$, minus the fraction of
mass that remains locked in the remnants from those same stars
\citep[see e.g.,][for empirical constraints on the
remnant mass fraction]{kalirai_etal08}. 
Here we will use mass loss rates tabulated from stellar evolution tracks to estimate the amount of mass lost by an 
SSP, but 
calculating mass loss from a main sequence turnoff
time \citep{raiteri_etal96} and the initial-final mass relation
\citep[from][]{kalirai_etal08} as in \cite{Agertz2010}, results in 
only small differences ($5-10\%$ less mass loss in the first $2\Gyr$) from
what is presented below.

\begin{figure*}[ht]
\begin{center}
\vspace{.5cm}
\resizebox{7in}{!}{ 
\includegraphics[width=0.5\linewidth]{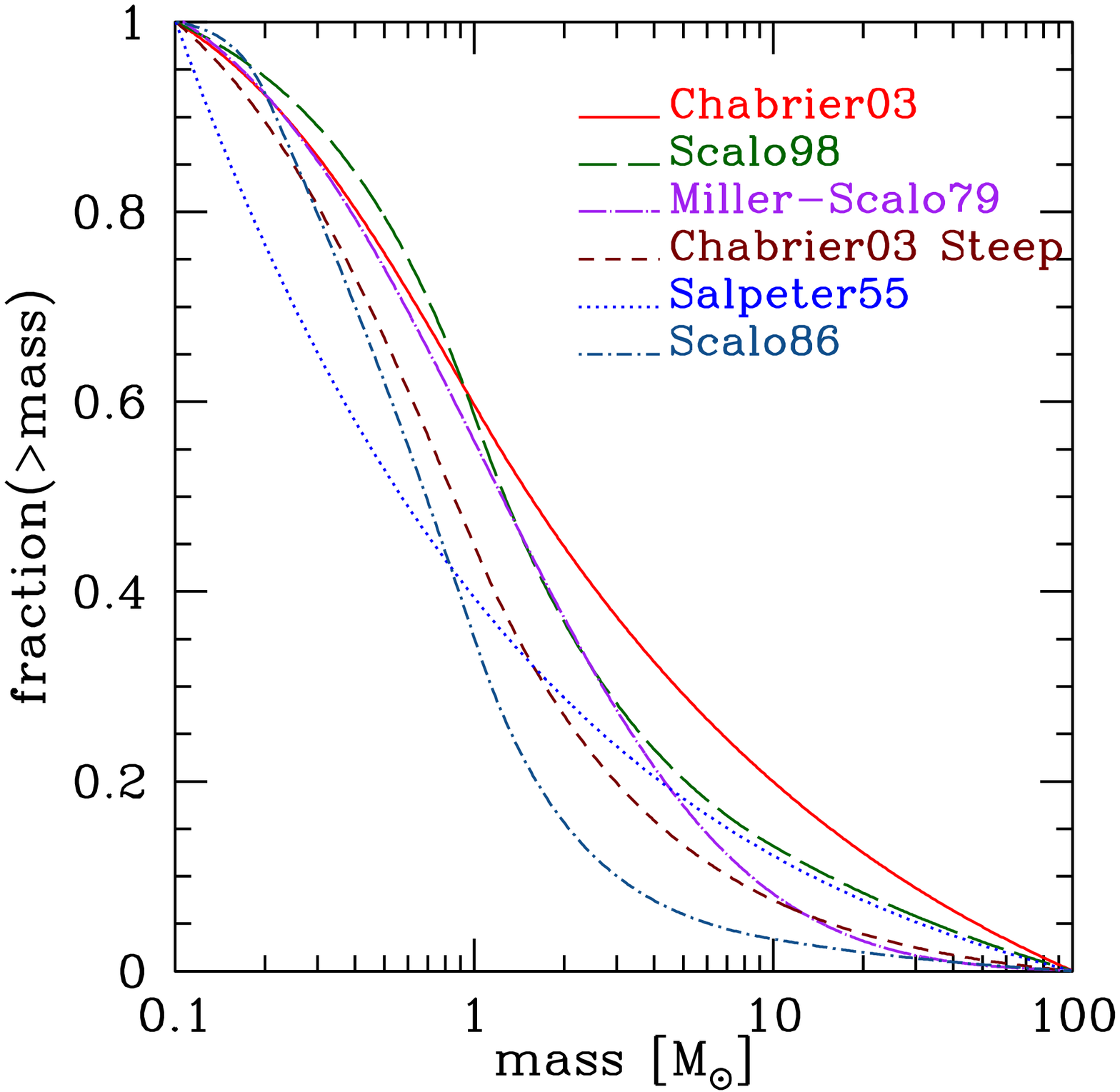}\hspace{0.5ex}
\includegraphics[width=0.5\linewidth]{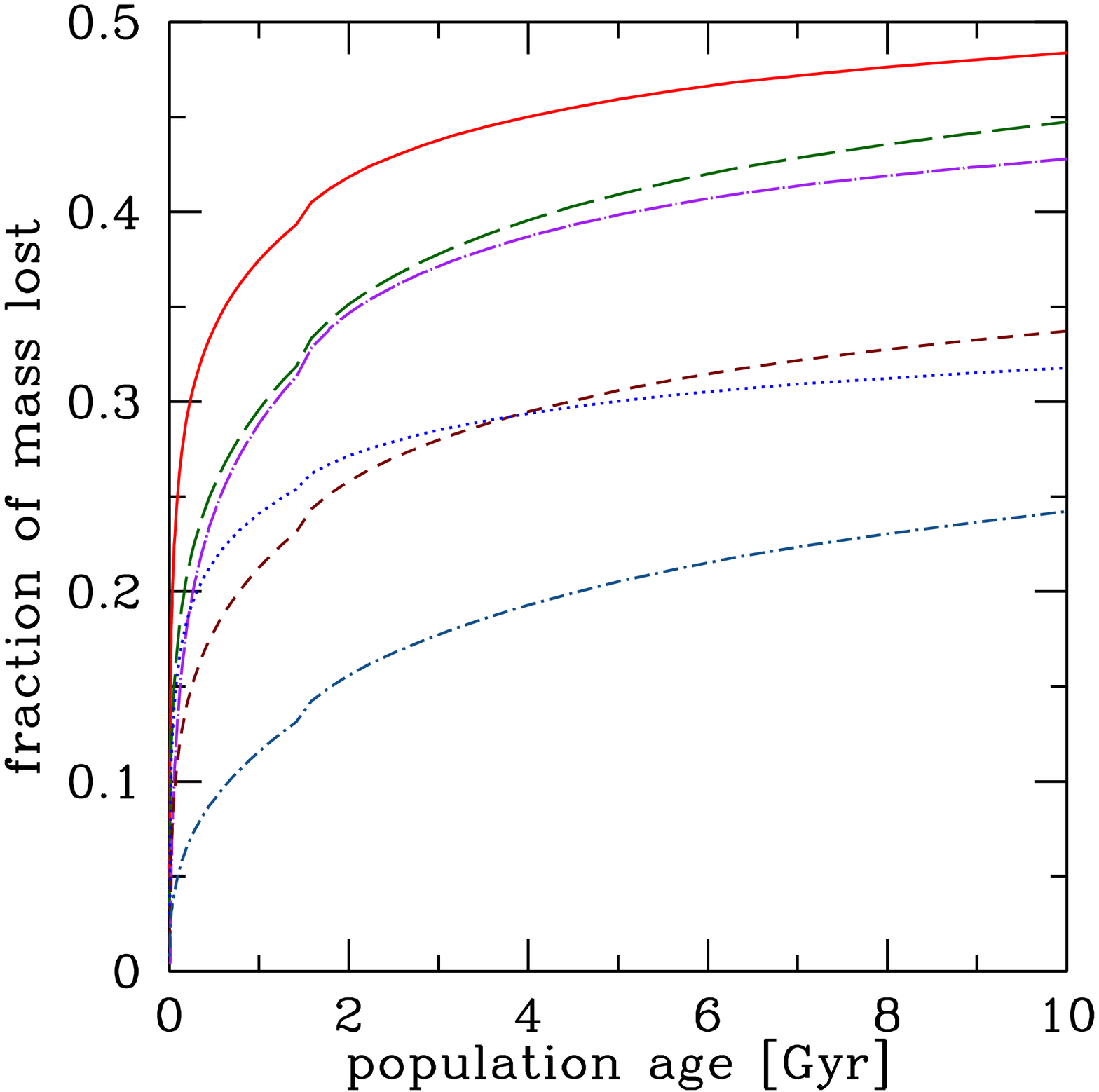} 
}
\end{center}
\caption{
\textit{Left panel}: The fraction of stellar mass in stars of mass $>m$ for a number of commonly used IMFs. 
\textit{Right panel}: The fraction of mass lost as a function of age for a single-age stellar population. Chabrier03 and Chabrier03 Steep were chosen to bracket the effect of likely IMF choices on mass loss rates in our study (see text for details). 
}
\label{fig:fmlimf} 
\vspace{.5cm}
\end{figure*}

The underlying IMF sets the fraction of stars with mass above $m_{\rm to}(t)$. 
A number of commonly used IMFs are parameterized in
table~\ref{table:fmlimf}, where we define $\xi$ to be the number of
stars per logarithmic mass interval. These parameterizations include
broken power laws for different mass intervals,
\begin{equation*}
\xi(\ln(m))\propto m^{-\Gamma}
\end{equation*}
and lognormal distributions,
\begin{equation*}
\xi(\ln(m))\propto \exp\left[ \frac{-(\log_{10}(m)-\log_{10}(m_c))^2}{2\sigma^2} \right].
\end{equation*}
Corresponding cumulative stellar fractions for different IMFs are
 plotted in the left panel of Figure~\ref{fig:fmlimf}, which shows
 that systematic differences between IMFs at $1M_{\odot}$, which
 corresponds to the main-sequence turnoff mass after the $10\Gyr$ ,
 are $\approx 20\%$. At larger masses, corresponding to younger
 population ages, the differences can be as large as a factor of
 $\approx 2-3$. 

The right hand panel of figure~\ref{fig:fmlimf} shows cumulative
 stellar mass loss as a function of time for SSPs
 with the same IMFs and of solar metallicity. These were calculated in FSPS \citep[][]{Conroy2010}
 using the Padova stellar evolution models\footnote{\url{http://stev.oapd.inaf.it/cgi-bin/cmd}} \citep[based on][]{Marigo2008,Marigo2007,Girardi2000} 
 
The figure shows
 that the majority of stellar mass loss occurs in the first two
 billion years, but there is a persistent slow rate of mass loss
 continuing to late times. The differences in mass loss for different
 IMFs after several billion years can be as large as a factor of two.
 However, for the range of IMFs considered to be likely for stellar
 populations in normal galaxies \citep[see, e.g.,][]{Kroupa2007} the
 differences are $\lesssim 30\%$.

The fraction of mass lost for each case is well fit by 
the functional form of \cite{Jungwiert2001}. Namely, the cumulative
fraction of mass lost by a stellar particle at a time $t$ since its
birth is given by
\begin{equation}
\label{eq:fml}
f_{\rm ml}(t) =C_0 \ln \left( \frac{t}{\lambda} + 1 \right)
\end{equation}
fit parameters $C_0$ and $\lambda$ are given in table~\ref{table:fmlimf}. 

Although the rates of mass loss in any particular phase of stellar
 evolution may be metal dependent and mass loss rate measurements for
 individual stars in that phase may exhibit factor of $\lesssim 4$
 scatter \citep[e.g.,][]{Mauron2010}, the important quantity here is
 the observed initial-final mass relation. This appears to be a weak
 function of metallicity, such that a $Z=0.1Z_\odot$ population will
 lock approximately the same amount of mass in remnants as a solar
 metallicity population (e.g., \citealp{kalirai_etal08},\citealp{Marigo2007}).
Uncertainty in the IMF is therefore the dominant source of uncertainty in estimating the amount of mass loss from a SSP.

\begin{table*}[!t]
\centering 
\caption{Mass loss and the IMF} 
\label{table:fmlimf}
\begin{tabular}{ l | c | c  | c  } 
\hline \hline
 & &\multicolumn{2}{ c }{Mass loss fits} \\[-.5ex]
\raisebox{1.5ex}{IMF} & \raisebox{1.5ex}{Functional form} 
 & $C_0 $ & $\lambda[\yr]  $
\\ [0.5ex]
\hline
&&&\\[-0.5ex] 
\cite{Chabrier2003} [C03]               &  $m<1\Msol$: log-normal ($\sigma=0.69$, $m_c=0.08$) & $0.046$ & $2.76\times10^5$ \\[0.5ex] 
(similar to \citealp{Kroupa2001})       &  $m>1\Msol$: power-law ($\Gamma=1.3$)  & &  \\[0.5ex]
\hline
&&&\\[-0.5ex] 
Steep \cite{Chabrier2003} [stpC03]      &  $m<1\Msol$: log-normal ($\sigma=0.69$, $m_c=0.08$) &$0.051$ & $1.33\times 10^7$\\[0.5ex]
(similar to \citealp{Kroupa1993})       &  $m>1\Msol$: power-law ($\Gamma=1.7$)  & &     \\[0.5ex]
\hline 
&&&\\[-0.5ex] 
                                        & $m<1\Msol$: $\Gamma_1=0.2$ & &                      \\[0.5ex]
\cite{Scalo1998} [S98]& $1\Msol<m<10\Msol$: $\Gamma_2=1.7$           & $0.062$ &$6.97\times10^6$\\[0.5ex]
                                        & $m>10\Msol$: $\Gamma_3=1.3$                  & &   \\[0.5ex]
\hline
&&&\\[-0.5ex] 
\cite{Miller1979}                       & log-normal ($\sigma=0.74$, $m_c=0.95$) & $0.058$&$6.04\times10^6$  \\[0.5ex]
\hline
&&&\\[-0.5ex] 
\cite{Salpeter1955}                     & $\Gamma=1.35$                         & $0.032$ & $5.13\times10^5$  \\
\hline
&&&\\[-0.5ex] 
                                   & $m<0.2: \Gamma_1=-2.6; 0.2<m<1\Msol$: $\Gamma_2=0.80$   & &\\ 
\cite{Scalo1986}$^a$               & $1\Msol<m<10\Msol$: $\Gamma_3=2.25$   &$0.055$ &$1.25\times10^8$ \\
                                   & $m>10\Msol$: $\Gamma_4=1.45$   & & \\
\hline                                                                                                                 
\hline
\end{tabular}

{$^a$\footnotesize{A fit from \cite{Mo2010} with an extension to $m<0.2$.}}

\end{table*}

To represent the plausible range of IMFs \citep[see recent review
by][for an extensive discussion of observational constraints on the
possible range of IMFs]{Bastian2010}, we adopt the commonly
used \citet[][hereafter C03]{Chabrier2003} IMF and its steep version,
stpC03. The latter has the same lognormal form as the C03 IMF below
$1M_\odot$, but a slope that is steeper than \cite{Salpeter1955} by
$0.35$ above $1M_\odot$. These choices bracket the range of IMFs that
are favored by observations\footnote{The C03 and stpC03 cumulative
mass functions vary by no more than $5\%$ from the \cite{Kroupa2001}
and \cite{Kroupa1993} mass functions respectively, meaning their mass loss rates are
very similar.}: they include the observed flattening of the mass
function at low masses \citep{Miller1979, Kroupa1993}, and also
encompass the range of high mass slopes that are allowed by
observations, when theoretical uncertainties in unresolved binary
stars \citep{Kroupa1991,Kroupa2007}, dynamical evolution in clusters
and other effects are considered \citep{Kroupa2003,Bastian2010}.
Although they are listed in our tables and shown for comparison, we
choose not to bracket the low mass loss case with either the original
power law \cite{Salpeter1955} IMF or the \cite{Scalo1986} IMF because
both are thought to predict too many low mass
stars \citep[$<1\Msol$][]{Miller1979,Kroupa1993,Scalo1998,Chabrier2001},
and over predict dynamical mass \citep{Bell2001,Cappellari2006}. Unless
otherwise noted, C03 will be used as the fiducial IMF in this study.

\section{The Star Formation and Mass Loss Histories of Galaxies}\label{sec:globalobservedsfh}

With mass loss rates for SSPs in hand, the
age distribution of stellar populations that are losing mass, i.e.
star formation histories, are needed to determine the relevance of the
galaxy-wide mass loss to the gas budget and for fueling continuing
star formation at low $z$.

To this end, we will use a model of SFHs derived from the observed
relation between SFR and stellar masses of galaxies at different
redshifts. The SFHs can be derived in this way because stellar mass in
disk galaxies is primarily the integral of the in-situ SFR modulo mass
loss. Given the observed evolution of the SFR-$M_{\ast}$ relation with
redshift for actively star forming galaxies, SFHs for galaxies of a
given current stellar mass, $M_{\ast 0}$, can be derived by starting
with $M_{\ast 0}$, integrating the stellar mass back in time and
determining the corresponding SFR at each epoch $z$ using the evolving
$M_{\ast}(z)$ and the SFR-$M_{\ast}$ relation at that epoch. Such a
procedure is justified because scatter in the SFR-$M_{\ast}$ relation
is relatively small: the distribution of $\SFR(M_{\ast})$ at a given
$z$ is approximately log-normal with a $1\sigma_{\ast}$ of $\approx
0.3$ dex \citep[][]{Noeske2007}. This implies that the galaxy
population spends $\approx 95\%$ of its time, and builds up $\approx
87\%$ of its stellar mass, during periods when SFRs are within a
factor of four ($<2\sigma_{\ast}$) of the median at a given epoch.
Nevertheless, we explore the effect of the scatter in the
SFR-$M_{\ast}$ relation on mass loss rates more thoroughly in
\S~\ref{sec:modelingscatter}. A more detailed discussion of average
star formation histories derived from measures of instantaneous star
formation rate will be presented in a separate paper \citep[][in
preparation]{Leitner2011}. Although SFHs of individual galaxies might
be determined more accurately by fitting population synthesis models
to galaxy spectra, the approach we adopt here should provide an
average estimate of the importance of stellar mass loss for galaxies
of a given stellar mass.

Implicit in this strategy for deriving SFHs is the assumption that
 mergers play no significant role in stellar build-up and that a
 galaxy's stellar mass changes solely because it forms stars or loses
 mass through stellar mass loss. Mergers have the potential to split
 $M_{\ast}$ as lookback time increases, thereby slightly increasing
 the SFR for the ensemble of progenitors compared to the values
 derived using our model. If mergers were an important factor in the
 growth of low-$z$ galaxy stellar mass, the SFR derived using our
 model would be underestimated resulting in galaxies forming
 more of their stellar mass at shorter lookback time. There
 is evidence, however, that the effect of mergers on the buildup of stellar mass
 is not large for most disk galaxies of $M_{\ast 0}\lesssim1 0^{11}\rm
 M_{\odot}$ at low redshifts. For example, both \citet[][see
 Figure~4]{Bell2007} and \cite{Drory2008} show that a scheme similar to ours reproduces the
 buildup of the stellar mass function of all galaxies with 
 only a minor contribution from mergers. Furthermore, \cite{Conroy2009} derive average SFHs
 for galaxies using the subhalo abundance matching method and show that
 mergers of galaxy halos cannot play a significant role in the evolution
 of the stellar mass function of galaxies at $z<2$.

\subsection{The Observed Evolution of the SFR-$M_{\ast}$ Relation} \label{sec:scaling}

\begin{figure}[!t]
\begin{center}
\vspace{.5cm}
\resizebox{3.3in}{!}{\includegraphics{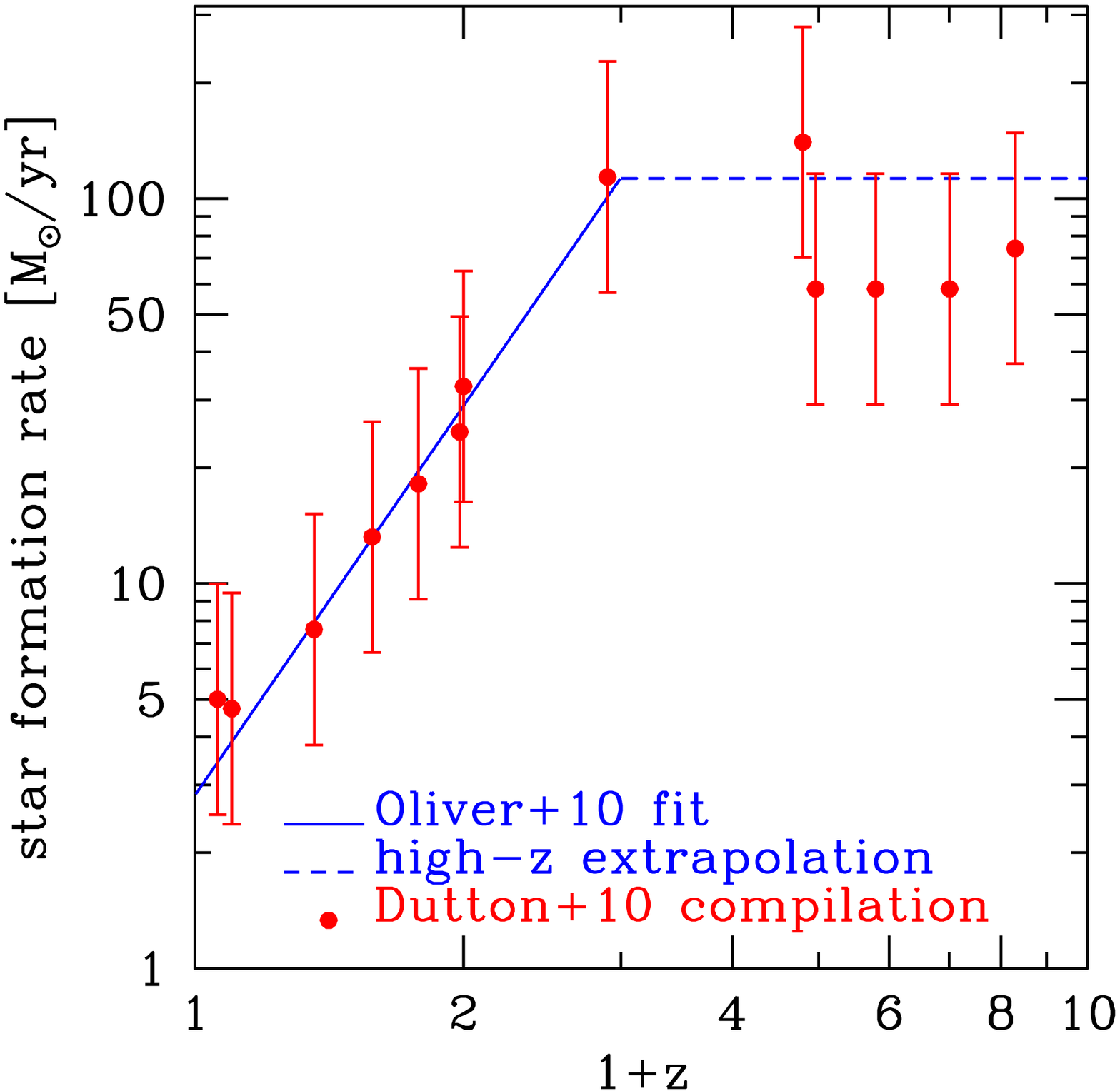}}
\end{center}
\caption{
Observations of the evolution of the normalization of the
 SFR-$M_{\ast}$ relation at the stellar mass
 $M_{\ast}=10^{10.75}\Msol$ with redshift for various data sets. The
 red points are from the \cite{Dutton2010} compilation of data
 from \cite{Gonzalez2010,Stark2009,Daddi2009,Daddi2007,Elbaz2007,Noeske2007,Salim2007}.
 The error bars show the estimated factor of two scatter of galaxies about
 $\psi(M_\ast,z)$. At $z\lesssim 2$, the blue line is the fit to the far infrared data of
 late-type star forming galaxies from \cite{Oliver2010}, which shows
 good agreement with previous studies. At $z>2$, we assume that
 normalization of the SFR-$M_{\ast}$ is constant (dashed line).}
\label{fig:SFRdata} 
\vspace{.5cm}
\end{figure}

A number of studies have explored the evolution and shape of the
SFR-$M_{\ast}$ relation at complementary UV and infrared wavelengths
\citep[][see the compilation of results by 
\citeauthor{Dutton2010} \citeyear{Dutton2010}]{Noeske2007,Elbaz2007,
Salim2007, Daddi2007,Daddi2009,Stark2009, Gonzalez2010}. Recently,
\cite{Oliver2010} performed a far-infrared stacking survey at $70$ and
$160\mic$, and included $100\mic$ and $160\mic$ observations from the
Herschel satellite \citep{Rodighiero2010}, which allowed them to probe
regions of the SED less affected by uncertainties in dust absorption.
\cite{Oliver2010} derived a fit to the median evolution and shape of
the SFR-$M_{\ast}$ relation of star forming galaxies in their sample.
The redshift evolution they derive is plotted as a solid line along
with the points from previous studies in Figure~\ref{fig:SFRdata}. The
figure shows a general consistency between different observations at
$z\lesssim 1$ (points have been scaled to $M_{\ast}=10^{10.75}\Msol$
according to the SFR-$M_{\ast}$ relation provided for each separate
data set in \citealp{Dutton2010}). Studies including quiescent
galaxies \cite[e.g.,][]{Zheng2007} also match the recent IR-derived
results \citep{Oliver2010}, indicating that systematic errors are
under control \citep[although see][for recent $L_{IR}$-SFR calibration
from Herschel]{Rodighiero2010}. In this study, we use the fit of
\cite{Oliver2010} at $z<2$.

At $z>1$ uncertainty in the normalization and slope of the
SFR-$M_{\ast}$ relation increases and there is a possibility that an
evolving IMF may affect the evolution \citep[][but
see \citealp{Bastian2010}]{Dave2008,Wilkins2008}. Fortunately, the
build-up of stellar mass at $z>2$ does not significantly affect
estimates of stellar mass loss at $z=0$ for late type galaxies, which
are the focus of our study. Furthermore, late-type galaxies with
$M_{\ast}<10^{11}\Msol$ appear to build up most of their stellar mass at
$z\lesssim 1$, so the evolution of star formation rates at high
redshifts is not very important (see figure~\ref{fig:avgsfh}). We
mimic the high redshift evolution with a redshift independent SFR that
forms a broken power law with $z_{\rm break}=2$.

Recent measurements of the slope $\beta$ in the relation
$\SFR~\propto~ M_{\ast}^{\beta+1}$ at fixed epoch, show $-0.5<\beta<0$
at $z<1$. Whether $\beta$ remains constant with
redshift \citep{Zheng2007}, evolves to a shallower
value \citep[e.g.,][]{Brinchmann2004, Noeske2007, Elbaz2007}, or
steepens \citep{Rodighiero2010,Erb2006,Oliver2010}, is unclear at
present. Here we assume a constant model with $\beta=-0.25$. The
resulting global scaling we use for $z<z_{\rm break}$ is,
\begin{equation}\label{eq:scaling}
\psi(M_{\ast},z) = A_0 \left( z+1 \right)^{\alpha} \left( \frac{M_*}{10^{10.75}\Msol} \right)^{1+\beta},\\
\end{equation} 
where we adopt the power law slope of $\alpha\approx 3.4$ for
late-type galaxies of $M_{\ast 0}=10^{10.75}\Msol$
\citep[][]{Oliver2010}. We label this fit to the median instantaneous
SFR measurements $\psi$ to distinguish it from the SFR that a specific galaxy
experiences as a function of redshift (its SFH). 
The normalization $A_0$ to $\psi(M_\ast,z)$ (plotted in
figure~\ref{fig:SFRdata} for $M_\ast=10^{11}M_\odot$) is
$2.8\Msol\yri$ \citep[][]{Oliver2010}. Uncertainty in $\alpha$ and
$\beta$ will be discussed further in section~\ref{sec:avgsfh}.

For $z_{\rm break}<z<6$ we use $\alpha=0$, and assume that all initial
stars in a galaxy form in a burst at $z=6$ (these stars 
constitute only a few percent of $M_{\ast 0}$).

\subsection{Calculating Star Formation and Mass Loss Histories}\label{sec:avgsfh}

The overall global mass loss rate (GMLR) of a
 galaxy is equal to its SFR convolved with the fractional mass loss
 rate $\dot{f}_\ml$ (whose integral is given by eq.~\ref{eq:fml}):

\begin{equation}
\label{eq:grr}
\GMLR(t) = \int_{t_0}^t \SFR(t') \dot{f}_{\rm ml}(t-t') dt'
\end{equation}

In our toy model, the evolution of the stellar mass of a galaxy is given by,
\begin{equation}
\dot{M}_{\ast}(t) = \psi(M_{\ast},z) - \GMLR(t)
\end{equation}
with boundaty condition $M_\ast(t=0)=M_{*0}$, where $\psi(M_{\ast},t)$ is the instantaneous SFR function from
eq.~\ref{eq:scaling}, which is a function of the {\it evolving} mass of the galaxy $M_\ast$. The SFH of the galaxy is then given by, 
\begin{equation}
\SFR(t) = \psi(M_{\ast}(t),z)
\end{equation}

\begin{figure*}[!t]
\begin{center}
\vspace{.5cm}
\resizebox{7in}{!}{
\includegraphics[]{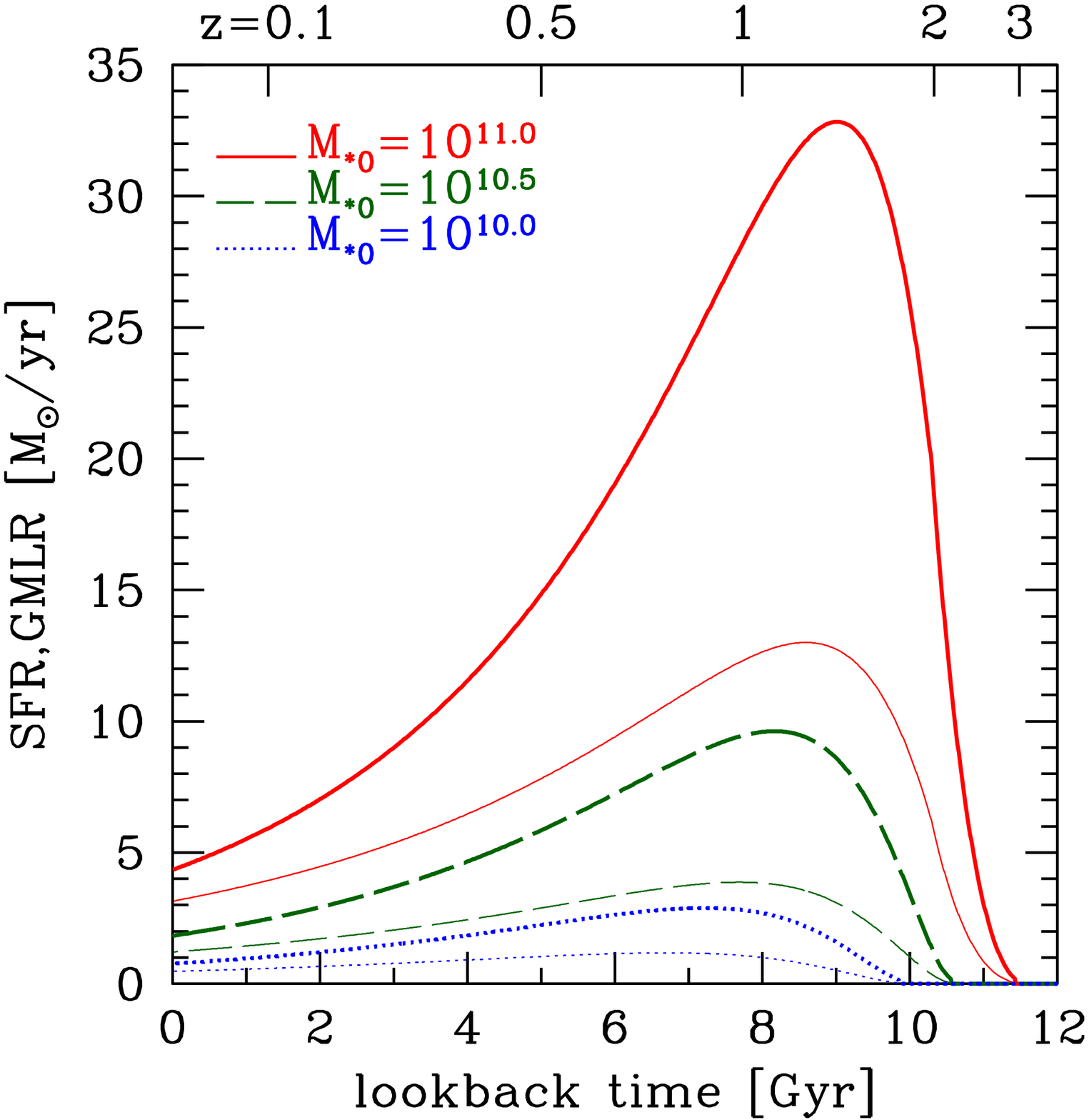}
\includegraphics[]{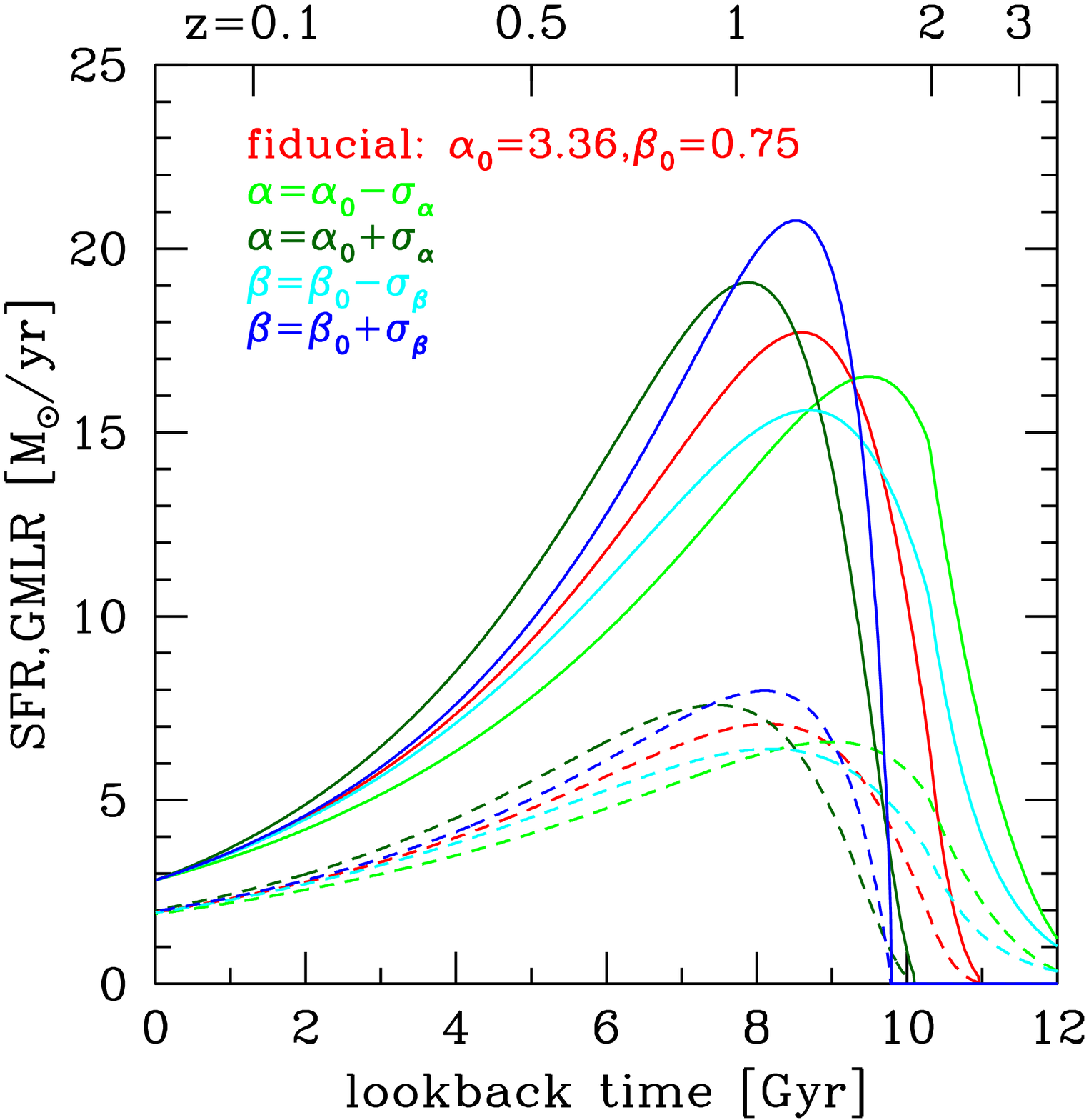}
}
\end{center}
\caption{ {\it Left Panel:}
 SFHs ({\it thick lines}) for galaxies of different $z=0$ stellar
mass, $M_{\ast 0}$: $10^{11.0}\Msol$ (red solid line),
$10^{10.5}\Msol$ (green dashed line), and $10^{10.0}\Msol$ (blue
dotted line). SFHs here are derived from the observed $\psi(M_{\ast},z)$
(see \S~\ref{sec:avgsfh} for details). The {\it thin lines} of the
corresponding color and line type show the global stellar mass loss
rates for these SFHs. {\it Right Panel:} SFHs ({\it solid lines})
and stellar mass loss rates ({\it dashed lines}) calculated for a
galaxy of $z=0$ stellar mass of $M_{\ast 0}=10^{10.75}\Msol$ with the
values of parameters of $\psi(M_{\ast},z)$, $\alpha$ and $\beta$ (see
eq.~\ref{eq:scaling}) varied by $1\sigma$ ($\sigma_{\alpha}\approx0.5$
and $\sigma_\beta\approx0.25$) from their average measured values
(lines of different color, with color correspondence indicated in the
legend).  Note that stellar mass loss rate at low $z$ is insensitive
to $1\sigma$ variation of the parameters of $\psi(M_\ast,z)$ at this level. }
\label{fig:avgsfh} 
\vspace{.5cm}
\end{figure*}

The $\GMLR(t)$ term introduces a complication in that it is defined 
as a convolution of the SFR of a galaxy with its fractional mass loss
rate up to time $t$, but the SFH of the galaxy is not, at first, known.
As a first guess, we can take $\GMLR(t)=0$, derive $M_{\ast}(t)$, plug the resulting SFH into eq.~\ref{eq:scaling} and iterate to convergence, 
\begin{equation}
\label{eq:sfhiter}
\begin{split}
\dot{M}_{\ast}^{(0)}(t) &=  \psi(M_{\ast}^{(0)},t) \\                      
\SFR^{(0)}(t) &= \psi(M_{\ast}^{(0)}(t),z)\\
\dot{M}_{\ast}^{(1)}(t) &=  \psi(M_{\ast}^{(1)},t) - \GMLR(\SFH^{(0)},t) \\ 
\SFR^{(1)}(t) &= \psi(M_{\ast}^{(1)}(t),z)\\
&  \vdots 
\end{split}
\end{equation}
Star formation and
stellar mass loss histories are plotted in Figure~\ref{fig:avgsfh}
for several representative values of $M_{{\ast}0}$ and the fixed
median SFR scaling relation discussed in \S~\ref{sec:scaling}.

In the right panel of Figure~\ref{fig:avgsfh}, we show the effect of
the uncertainty in the parameters $\alpha$
($\sigma_{\alpha}\approx\pm0.5$) and $\beta$
($\sigma_{\beta}\approx\pm0.25$) on the SFH and the mass loss
rates \citep[based on estimates by][and dispersion in the literature
discussed above]{Oliver2010}.
 The observational $1\sigma$ uncertainty in the slope and evolution of the SFR-$M_{\ast}$ relation 
only translates into uncertainty in the mass loss rate of $<5\%$ at $z=0$, although the
differences can be relevant to the stellar mass loss history.

\subsection{The Impact of Scatter in the SFR-$M_{\ast}$ Relation}\label{sec:modelingscatter}

Although the observed scatter in the SFR-$M_{\ast}$ relation is
relatively small, deviations from the median relation can be
significant for some galaxies and may thus affect conclusions about
the importance of stellar mass loss. In this section, we will
therefore consider the effects of populating the scatter in
SFR-$M_{\ast}$ relation, with a range of possible SFH scenarios. We
parameterize the scatter around the star formation rate function $\psi(M_\ast,z)$
that an individual galaxy experiences, with the time scale,
$\tau_\Delta$, over which deviations around the median relation occur,
and by the magnitude of scatter, $\sigma_\ast$. We assume that the
scatter is described by a log-normal function with the median
described by the relation in equation~\ref{eq:scaling}. We thus
multiply $\psi(M_\ast,z)$ at a given redshift by by
$N(t,\tau_\Delta,\sigma_\ast)$ -- a log-normal random number sampled
every $\tau_\Delta$ for $t>\tau_\Delta$. To force the model to trend
toward the assigned $SFR(z=0)$, we set
$N(t,\tau_\Delta,\sigma_\ast)=SFR(z=0)/A_0$ for lookback times
$<\tau_\Delta$.

\begin{figure*}[!t]
\begin{center}
\vspace{.5cm}
\resizebox{7in}{!}{
\includegraphics[]{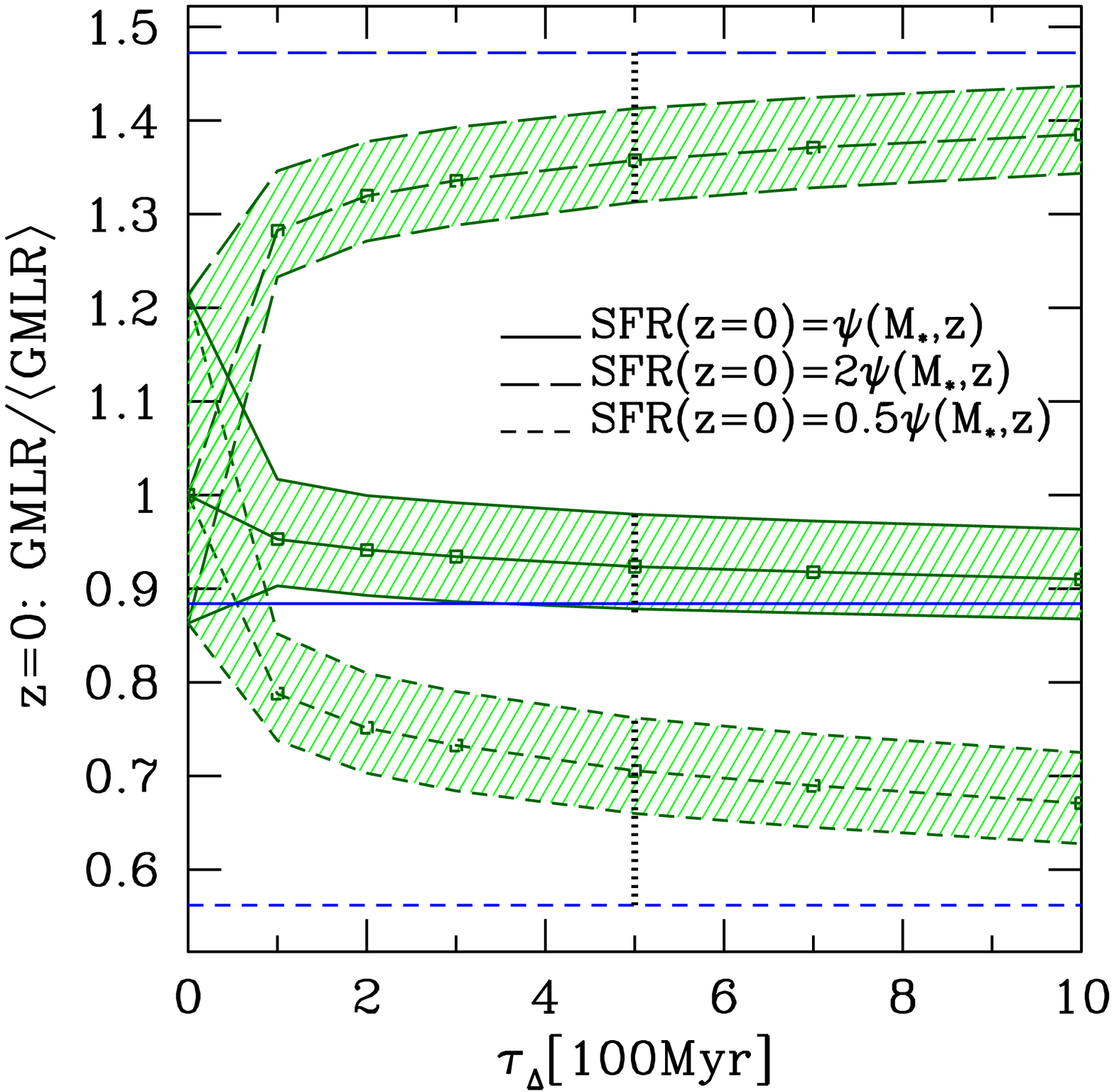}
\includegraphics[]{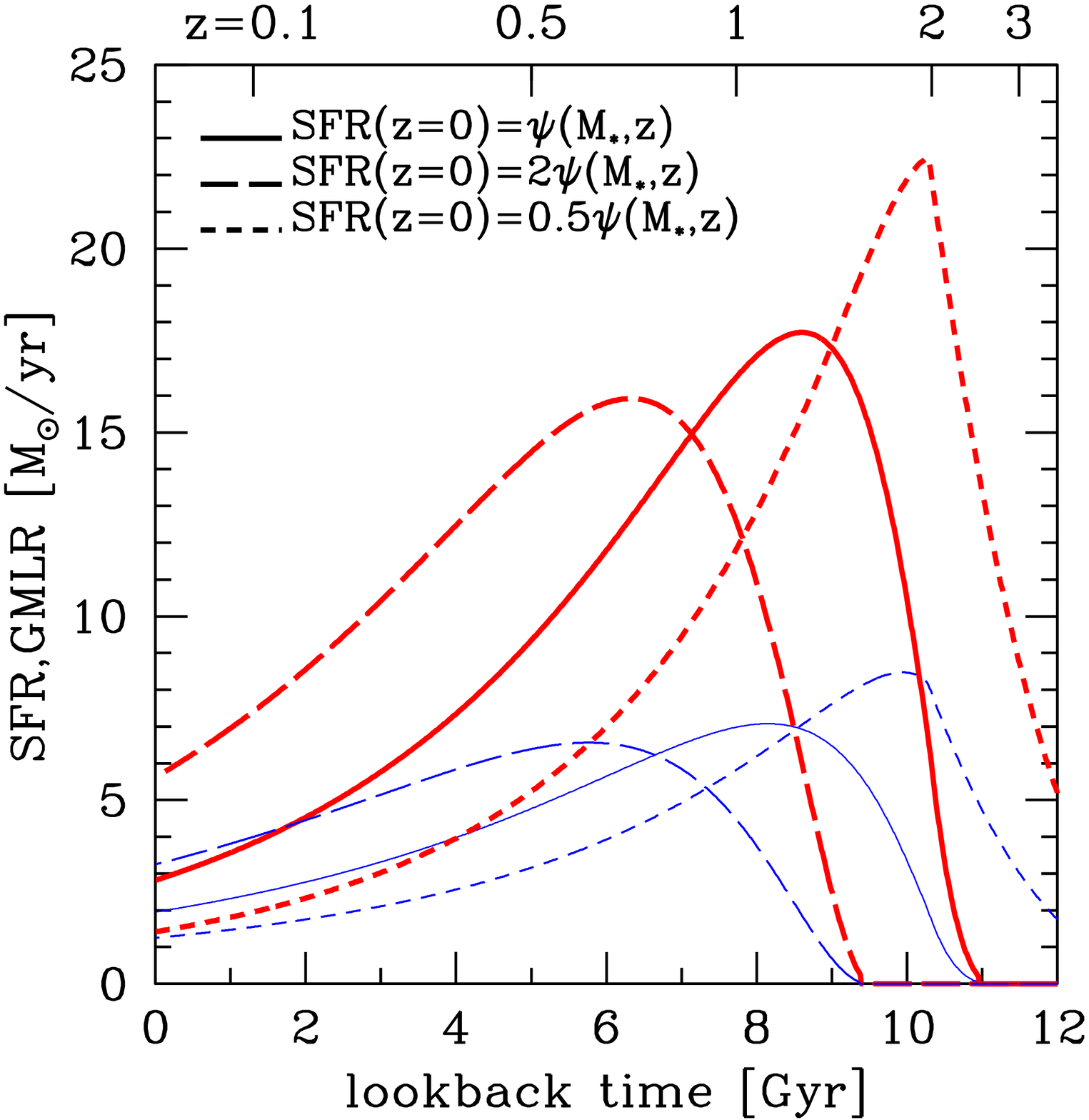}
}
\end{center}
\caption{
\textit{Left Panel:} 
The $z=0$ stellar mass loss rate for galaxies of
$M_{\ast0}=10^{10.75}\Msol$ as a function of $\tau_{\Delta}$
normalized to the population's average mass loss rate in a scenario
where galaxies draw their SFR randomly from a lognormal distribution
about $\psi(M_\ast,z)$ at all times (i.e., $\tau_{\Delta}=0$, see text for details). Solid,
short dashed, and long dashed lines correspond to galaxies with
present day SFR that are at the median ($A_0$), 16th ($0.5A_0$) and
84th ($2A_0$) percentile in the scatter about $\psi(M_\ast,z)$ at
$M_{\ast0}=10^{10.75}\Msol$. Green regions
encompass the $1\sigma_\Delta$ scatter for a population about their
corresponding relative mass loss rate, while blue lines show the
$\tau_\Delta=14\Gyr$ limit for each present day SFR. The
$\tau_\Delta=14\Gyr$ cases are systematically lower for each present
day star SFR because the mean of a lognormal distribution is offset
from the median. We take $\tau_{\Delta}>500\Myr$ so the difference
between the mass loss rate with $\tau_{\Delta}=500\Myr$ and $14\Gyr$,
noted with vertical black dotted lines, marks the uncertainty in the
mass loss model.
\textit{Right Panel}: 
Star formation (thick red) and stellar mass loss (thin blue) histories
at $M_{\ast 0}=10^{10.75}\Msol$ for $\tau_{\Delta}=14\Gyr$. Line
styles are the same as on the left. The model shows that stellar mass
loss can provide a larger fraction of the fuel required to maintain
star formation in galaxies with low $\SFR(z=0)$. In all cases, the
global stellar mass loss rate is a significant fraction of the present
day SFR.}
\label{fig:scatter} 
\vspace{.5cm}
\end{figure*}

The effect of scatter is illustrated in Figure~\ref{fig:scatter},
which shows relative global stellar mass loss rates as a function of
assumed $\tau_\Delta$ for model galaxies with
$M_{\ast}=10^{10.75}\Msol$ (the stellar mass of the Milky Way) and
$\SFR(z=0)$ corresponding to the median SFR at that mass, as well as
star formation rates at $\pm1\sigma_{\ast}$ from the median. Mass loss
rates are normalized by the population-average rate in the
$\tau_\Delta=0$ case. The overall amplitude of each line in the
vertical direction therefore indicates the factor by which the mass
loss rate is biased with respect to the case wherein galaxies, at any
instant, draw their SFR randomly from the lognormal population-wide
probability distribution function of SFRs. The shaded regions covers
the area between the $16\%$ and $84\%$ outliers from the median mass
loss rate at a given $\{\SFR(z=0),\tau_\Delta\}$.

The shaded region in the figure shows that that scatter in the
 SFR-$M_{\ast}$ relation results in small scatter in the mass loss
 rate for a given galaxy. The different present day SFRs also result
 in an overall change of the mass loss rate, although this effect
 disappears for very short $\tau_\Delta$.

 Galaxy-wide stochasticity would presumably have a catalyst, such as a
 merger, tidal interaction, or rapid increase in gas accretion rate.
 These processes should operate on the crossing time of the halo
 ($\sim 1$~Gyr) so we adopt $\tau_{\Delta}=500\Myr$ as a fiducial
 minimum value. We do not enforce a maximum $\tau_\Delta$, thus we use
 $\tau_{\Delta}=14\Gyr$ for the scenario in which the scatter around
 the median relation for the disk population is not generated by
 stochasticity in individual galaxies, but is instead caused by
 persistent environmental effects; these effects would be different
 for different galaxies in the population, but similar over time for a
 given galaxy. In other words, in such a scenario, a galaxy that has
 a SFR below $\psi(M_\ast)$ at some high redshift will maintain a low
 SFR with respect to $\psi(M_\ast,z)$ at all $z$. The overall uncertainty
 in the global mass loss rates for a given mass loss model,
 encompassing both model dependence and the $\pm1\sigma_\Delta$
 outliers from stochastic variation with $\tau_{\Delta}=500\Myr$, can
 then be quantified as the GMLRs encompassing these two regions. This
 uncertainty is marked by the black vertical dotted lines for each
 $SFR(z=0)$ in Figure~\ref{fig:scatter}, and this method will be used
 to describe model uncertainty in the results below.

The star formation and stellar mass loss histories for model galaxies
with $\tau_{\Delta}=14\Gyr$ are plotted in the right panel of
Figure~\ref{fig:scatter} for the same representative range of present
day SFRs. The lower $\SFR(z=0)$ case results in less star formation at
late times, and hence more star formation at high redshifts, as the
total stellar mass is kept constant. This push-back also results in
less stellar mass loss at low $z$, since a population loses more of
its mass when it is younger. However, the decrease in mass loss rate
is not proportional to the decrease in $\SFR(z=0)$, as is assumed by
the instantaneous \textit{recycling} approximation. This is because
old populations continue to shed mass. In fact, the mass loss and star
formation lines are very close to each other for the low $\SFR(z=0)$
case in Figure~\ref{fig:scatter}, indicating that stellar mass loss
could be responsible for fueling almost all of the star formation in
such a galaxy. Conversely, in galaxies with a higher current SFR for a
given stellar mass loss provides a smaller fraction of the fuel
required to maintain the star formation at the same level.

\section{Implications of Stellar Mass Loss for the Gas Supply in Galaxies}\label{sec:massloss_history}

\subsection{$z=0$ Stellar Mass Loss}

\begin{figure*}[!t]
\begin{center}
\vspace{.5cm}
\resizebox{7in}{!}{
\includegraphics[]{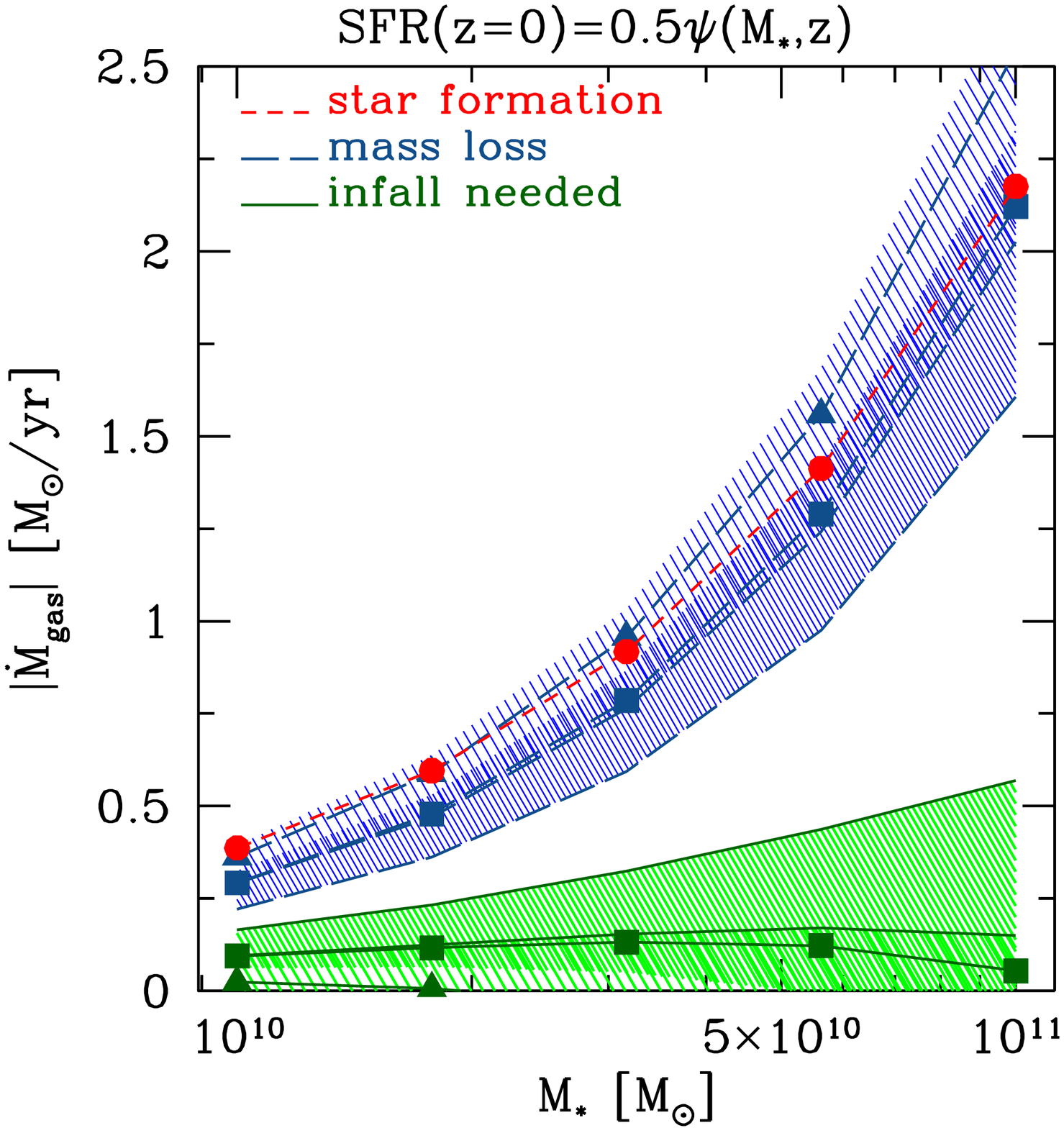}\hspace{0.0ex}
\includegraphics[]{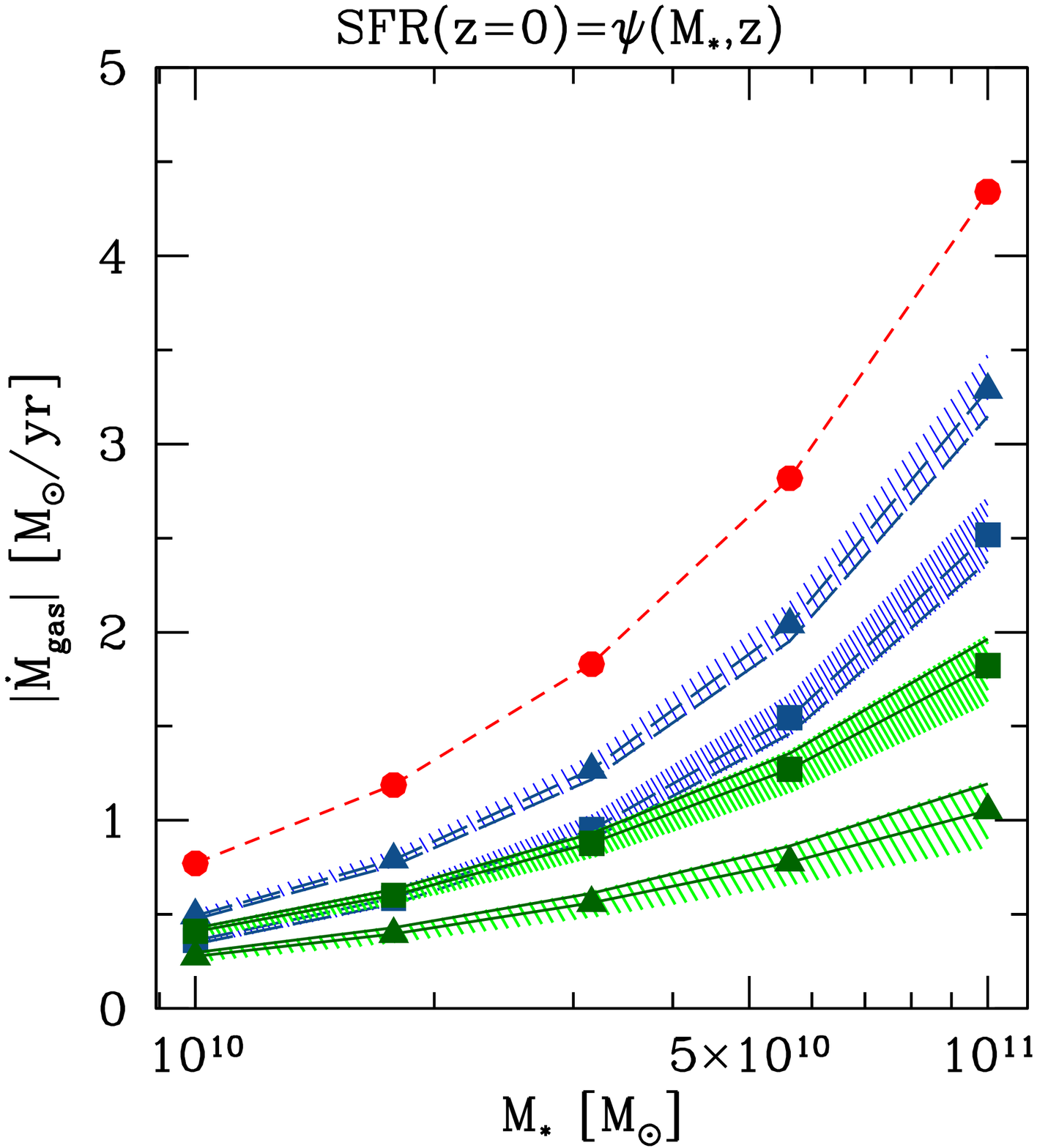}\hspace{0.0ex}
\includegraphics[]{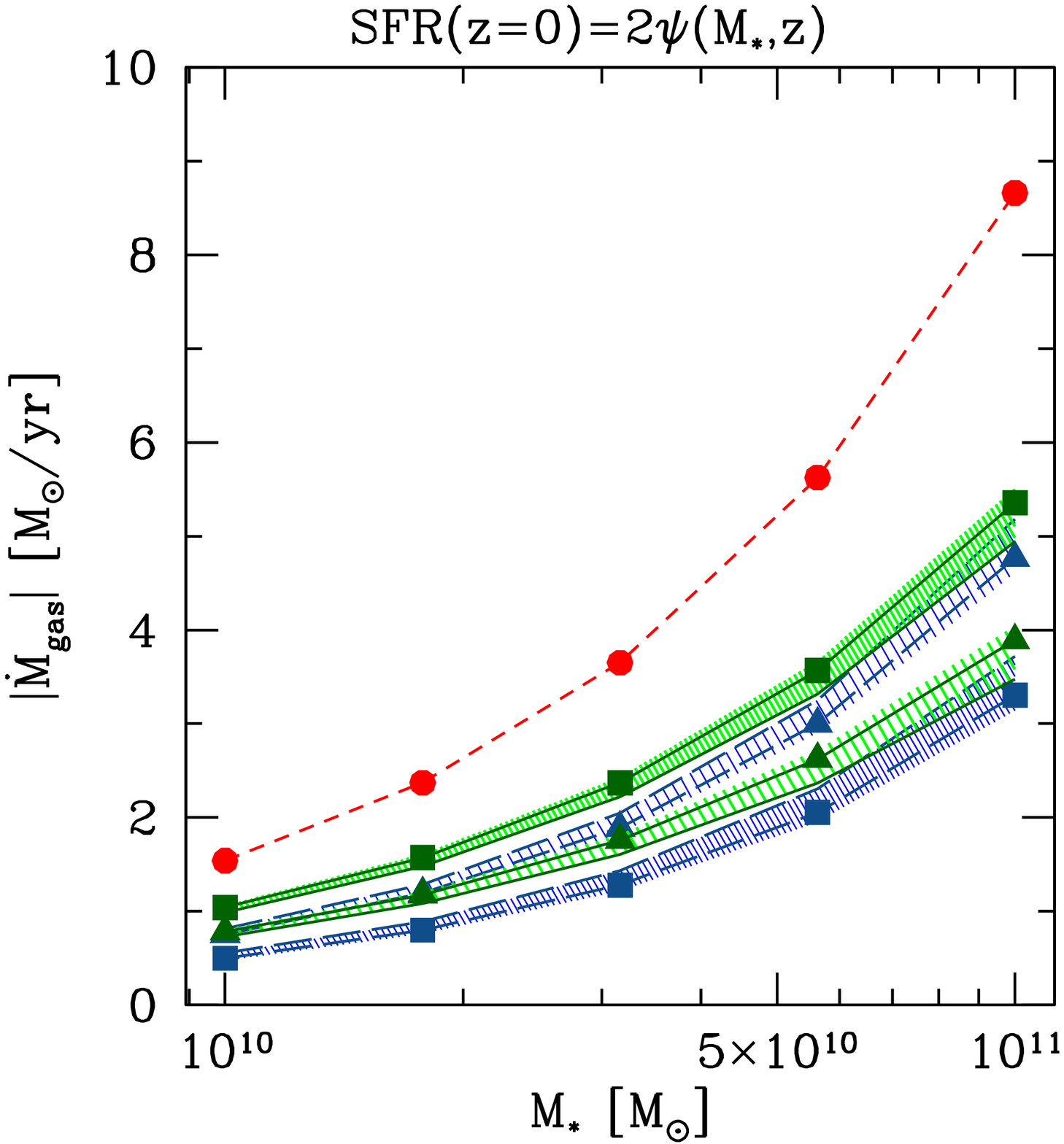}
}
\end{center}
\caption{
Average star formation rate (red dashed lines) and global stellar mass
loss rate (blue dashed lines) as a function of stellar mass. The green
solid lines show the difference between SFR and stellar mass loss
rate. Lines without points are calculated assuming no stochastic
scatter (i.e., $\tau_\Delta=14\Gyr$), and lines with points are for
the models with scatter assuming $\tau_\Delta=500\Myr$ timescale with
a C03 (triangles), and stpC03 (squares) IMFs. Shaded regions represent
modeling uncertainty for the C03 (lightly shaded) and stpC03 (densely
shaded) IMFs, which includes plausible range of stochasticity in the
SFR-$M_{\ast}$ relation. The central panel is for median present day
star formers, while the left and right panels are for galaxies with
star formation rates higher and lower than the median by
$1\sigma_{\ast}$, respectively. Note that in all panels stellar mass
loss rate is a significant fraction of the SFR and is comparable to
the SFR in the left panel (i.e., low SFR case). }
\label{fig:rep_z01}
\vspace{.5cm}
\end{figure*}

Present day stellar mass loss rates for galaxies, calculated assuming
both C03 and stpC03 IMFs (see \S~\ref{sec:modelingmassloss}) are
plotted in Figure~\ref{fig:rep_z01} as a function of galaxy stellar
mass. The additional infall that is required to match the consumption
of gas by star formation is the difference between the SFR and stellar
mass loss; it is shown by the hatched bands. The width of the shaded
regions corresponds to uncertainty about the C03 and stpC03 IMFs
caused by stochasticity (with $\tau_\Delta>500\Myr$) that can comprise
the scatter in the SFR-$M_{\ast}$ relation, as discussed
in \S~\ref{sec:modelingscatter}. Note that this uncertainty is
relatively small and does not affect our conclusions about the
importance of stellar mass loss.

The middle panel of Figure~\ref{fig:rep_z01} shows that stellar mass
loss dominates over infall across the entire plotted stellar mass
range. For a stpC03 IMF, stellar mass loss still makes up a
significant portion of the star formation but, on average, infall
should comprise $45-60\%$ of the star formation rate.

The side panels of Figure~\ref{fig:rep_z01} show the amount of stellar
mass loss for galaxies with present day star formation rates
$\pm1\sigma_\ast$ (i.e., a factor of two) away from the median
SFR-$M_\ast$ relation. The left panel shows that for galaxies with SFR
below the median, stellar mass loss dominates over infall for both C03
and stpC03 IMFs. The right panel shows that galaxies with SFR higher
than average require more significant infall to maintain their SFR.
For the stpC03 IMF infall should contribute about $60\%$ of the gas to
maintain current SFR, whereas for the C03 IMF, stellar mass loss still
dominates, but infall must make up for about $40-50\%$ of the gas
required to maintain the SFR.

\subsection{Recycling Epoch}\label{sec:recyclingepoch}

\begin{figure}[!t]
\begin{center}
\resizebox{3.3in}{!}{\includegraphics[]{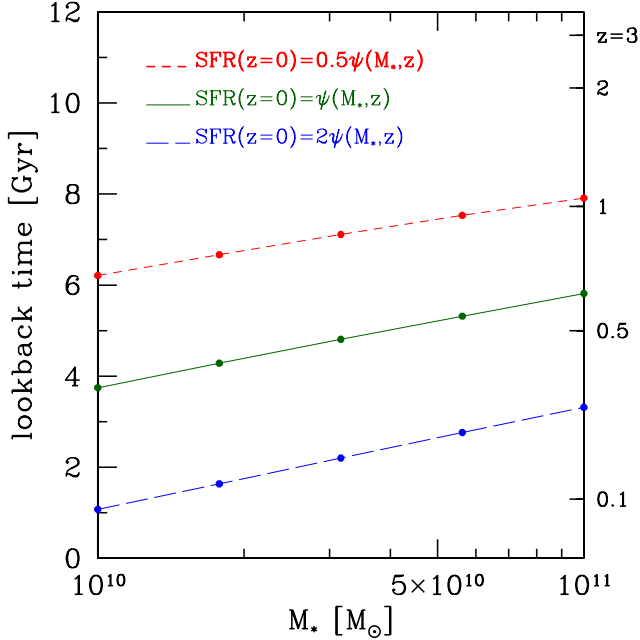}}
\end{center}
\caption{
The lookback time at which the global stellar mass loss rates become
larger than half the star formation rate, assuming a C03 IMF. The
short (long) dashed line show such time for galaxies in the 16th
(84th) percentile in $\SFR(z=0)$, while the solid lines show results
for the median SFR-$M_\ast$ relation. }
\label{fig:recyclingepoch} 
\vspace{.5cm}
\end{figure}

The epoch at which the stellar mass loss rate becomes larger than half of
 star formation rate is plotted in Figure~\ref{fig:recyclingepoch}.
 This period can be thought of as the {\it recycling epoch}, when
 stellar mass loss provides most of the new gas needed to replenish what is consumed by
 star formation. The figure shows that the recycling epoch
 occurs at $z\approx 0.7-0.9$, $z\approx 0.3-0.6$ and $\approx
 0.05-0.2$ for the galaxies with respectively systematically low, median, and high
 star formation rates relative to $\psi(M_\ast,z)$.
  The recycling epoch is sensitive to the
 assumed IMF; for example, assuming stpC03 IMF instead of C03, galaxies that follow $\psi(M_\ast,z)$ 
 enter the recycling epoch at $z\approx 0.1$.

\subsection{The Gas and Star Formation Budget of Nearby Galaxies}\label{sec:nearby}

It is only possible to estimate the cold gas accretion rate for a
handful of galaxies. Nevertheless, available estimates are all
considerably smaller than the current star formation rates of host
galaxies, a fact that has been used to raise concerns of a sizable gas
deficit
\citep{Fraternali2010,Fraternali2009,Sancisi2008,Fraternali2008,Putman2009,Marinacci2010}.
It is thus interesting to evaluate the potential contribution of
stellar mass loss.

\begin{figure*}[!t]
\begin{center}
\resizebox{7in}{!}{\includegraphics[]{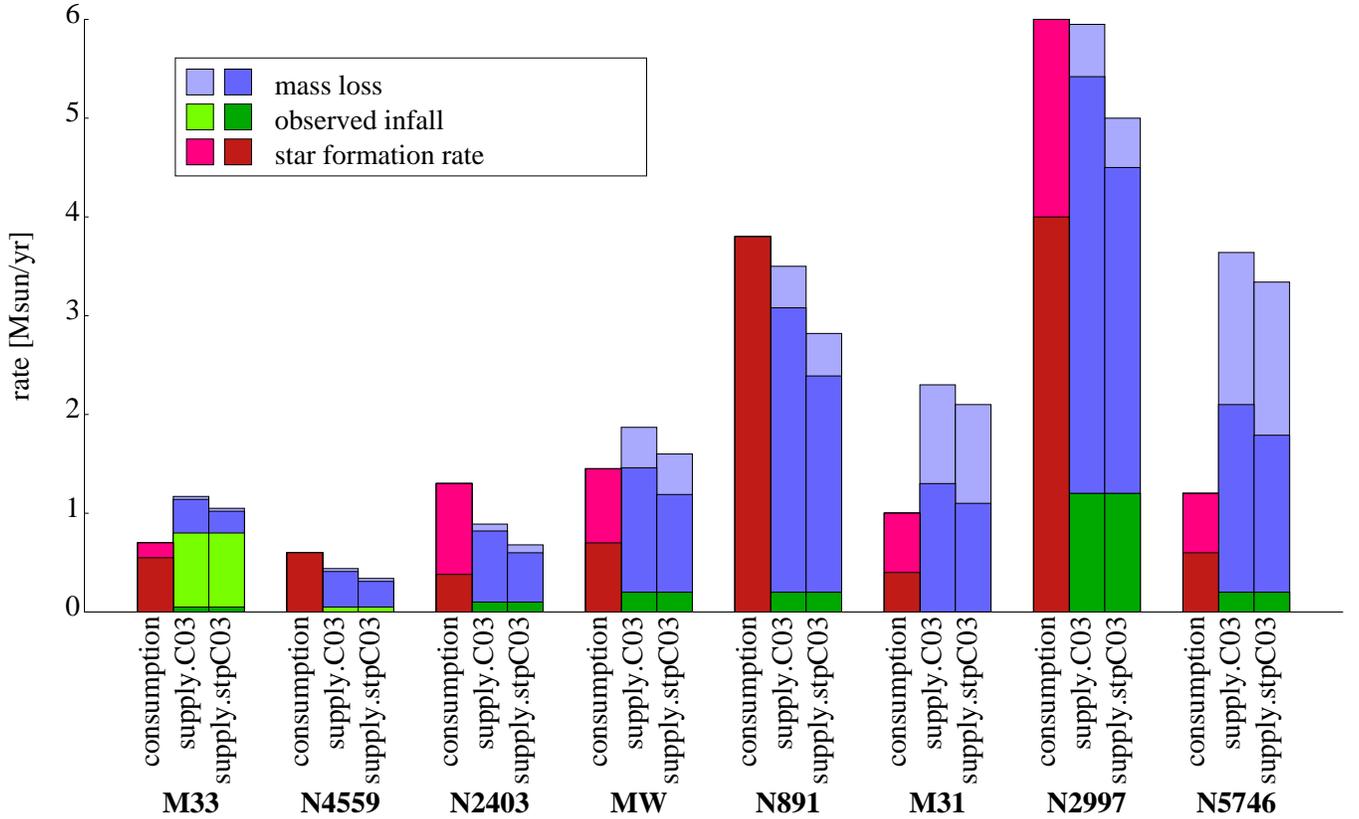}}
\end{center}
\caption{
The gas budget for each galaxy labeled on the bottom axis. Bars on the
left of each galaxy denote the observed gas consumption from star
formation (dark red and magenta). The two sets of bars on the right
denote gas supplied from observed infall of new gas (green) and
stellar mass loss (blue). The two gas supply rate columns are labeled,
below, by the C03 and stpC03 IMFs on which they are based. The lighter
shades of each quantity indicate the space between low and high
estimates of that quantity. For stellar mass loss, these are based on
the uncertainty estimates from this study; for star formation and
infall, they are based on the size of cited error bars and scatter
between different measurements. Specific values, references and notes
are given in Table~\ref{table:gasbudget}. }
\label{fig:gasbudget}
\end{figure*}

The gas budgets for the galaxies with available estimates of gas
infall rates are presented in Table~\ref{table:gasbudget} and
Figure~\ref{fig:gasbudget}, including the expected contribution
from stellar mass loss. To estimate the mass loss rate we use
observational constraints on $\SFR(z=0)$ and $M_{\ast 0}$ as input and
calculate present day stellar mass loss rates through modeled SFHs, as
described above. For galaxies with a range of $\SFR(z=0)$ quoted in
the table, we use the high value to calculate the mass loss
rates. This choice is conservative, because mass loss will make up a
smaller fraction of the gas supply in galaxies with higher SFR at
fixed stellar mass, as was shown above (see, e.g.,
Fig.~\ref{fig:scatter} right panel). Ranges quoted for global stellar
mass loss rates incorporate the scatter and modelling uncertainty that
was quantified in \S~\ref{sec:modelingscatter}.

\begin{table*}[!t]
\begin{center}
\caption{Galaxy properties and the gas budget.}\label{table:gasbudget}
\begin{tabular}{l c c c c c c c c} 
\hline\hline\noalign{\smallskip}
&& & &&& \multicolumn{2}{ c }{Stellar Mass Loss Rates} & \\[-.5ex] 
Galaxy&$M_{*0}$$^a$  & $\rm{v_{flat}}$  &median$(\SFR_0)$ &$\SFR_0$$^a$   &Accretion rate$^b$ & C03 & stpC03 & References \\
      &\footnotesize{($10^9\Msol$)}
      &\footnotesize{($\kms$)}
      &\footnotesize{($\Msol\yri$)}
      &\footnotesize{($\Msol\yri$)}
      &\footnotesize{($\Msol\yri$)}
      &\footnotesize{($\Msol\yri$)}
      &\footnotesize{($\Msol\yri$)}
      & \footnotesize{[$M_*0$][$\SFR_0$][Accr.]}\\
\noalign{\smallskip}
M\,33          \dotfill   & 4.5   &104&0.42& $0.55-0.7$                               &  $0.05-0.8$$^c$   &  0.34-0.37 &  0.22-0.25 & [1][2,3][4]\\
NGC\,4559  \dotfill   & 6.8   &123&0.58& 0.6                                          &  $<0.05$                               &  0.36-0.39 &  0.26-0.29 & [5][6][7]\\
NGC\,2403  \dotfill   & 5-14 &134&1.0& 0.38-1.3                                    &  0.1                                       &  0.72-0.79 &  0.50-0.58 & [8,9][10,11][12]\\
Milky Way  \dotfill   & 55    &220&2.8& $0.7-1.45$                                 &  0.2                                       &  1.26-1.67 &  0.99-1.40 & [13][14,15][16]\\  
NGC\,891   \dotfill   & 100   &230&4.3& 3.8                                           &  $0.2$$^d$          &  2.88-3.30 &  2.19-2.62 & [17][18][19]\\
M\,31          \dotfill   & 100   &260&4.3& $0.4-1.0$    &  0.0                                        &  1.30-2.30 &  1.10-2.10                 &[20][21,22,23][24]\\     
NGC\,2997  \dotfill   & 160   &226&6.2& $5.0\pm1$                             &  1.2                                        &  4.22-4.75 &  3.30-3.80   &[25][26][27]\\
NGC\,5746  \dotfill   & 160   &320&6.2& $0.8\pm0.2-1.2$                     &  $0.2$                                    &  1.90-3.44 &  1.59-3.14    & [28][29,30][31]\\
\hline
\end{tabular}
   \label{tExtra}
\end{center}
{\footnotesize 
$^a$Systematic and random errors in stellar mass and SFR are likely
$\lsim0.3\rm{ dex}$ \citep{Mo2010}, but difference between stellar
masses and SFR derived with C03 and stpC03 should be $<10\%$
(approximately the difference between the \citealp{Kroupa1993}
and \citealp{Kroupa2001} IMFs found by \citealp{Borch2006}).
$^b$Accretion rates are probably lower limits to cold-gas accretion,
and do not include helium or ionized gas fractions unless noted.
$^c$Infall of surrounding gas clouds at $100\kms$ is assumed, the low
(high) value excludes (includes) an ionization correction; $^d$If all
HI in the halo of NGC\,891 were to fall toward the disk it would
supply $\sim30\Msol\yri$;
\vspace{0.1cm}

\textbf{References:}
[1] \cite{Corbelli2003}; 
[2] \cite{Magrini2007}; 
[3] \cite{Blitz2006}; 
[4] \cite{Grossi2008}; 
[5] \cite{Barbieri2005}, from $M_*/L_k$; 
[6] \cite{Fraternali2010}; 
[7] \cite{Barbieri2005}; 
[8] \cite{Leroy2008}; 
[9] \cite{deBlok1996}; 
[10] \cite{Leroy2008}; 
[11] \cite{Kennicutt2003}; 
[12] \cite{Fraternali2002}; 
[13] \cite{Flynn2006}; 
[14] \cite{Robitaille2010}; 
[15] \cite{Murray2010}; 
[16] \cite{Wakker2007}; 
[17] \cite{Rhode2010}; 
[18] \cite{Popescu2004}; 
[19] \cite{Oosterloo2007}; 
[20] \cite{Chemin2009}; 
[21] \cite{Walterbos1994}; 
[22] \cite{Barmby2006}; 
[23] \cite{Williams2003}; 
[24] \cite{Chemin2009}; 
[25] \cite{Louise1983}; 
[26] \cite{Hess2009}; %
[27] \cite{Hess2009}; 
[28] \cite{Rasmussen2009}; 
[29] \cite{Rasmussen2006}; 
[30] \cite{Pedersen2006}; 
[31] \cite{Rand2008}; 
}
\end{table*}

The table and figure show that total gas supply rates (stellar mass
loss $+$ gas infall) are close to star formation rates.
Conservatively, the factor of several discrepancy between star
formation and cold-HI gas supply \citep[e.g.,][]{Fraternali2009}
should reduce to a factor of two discrepancy for some galaxies, and to
zero or a surplus in others.

Furthermore, a less conservative accounting of infall may include
factor of two or larger corrections to infall rates from (1) a $33\%$
correction for the helium fraction, and (2) the fact that neutral gas
only accounts for a fraction of the infalling cloud mass. The neutral
fraction in the HVC clouds is not very well constrained at present
with observational estimates ranging from $\sim 10-20\%$ for HVCs that
are distant \citep{Maloney2003} to $\sim 50\%$ for the clouds near the
disk \citep[][]{Wakker2007,hill_etal09}. Despite the uncertainties, it
is clear that ionized gas constitutes a non-negligible mass fraction
of the halo clouds \citep{sembach_etal03} and so the accretion rate
observed in HI can only be a lower limit. There could also be
additional possible sources of fresh gas, such as infall from gas rich
satellites ($0.1-0.2\Msol\yri$) and accretion that is lost in
confusion noise (esp. for NGC 891), as discussed
by \cite{Sancisi2008}. Observations of the Milky Way suggest that
$0.2\Msol\yri$ of additional infall may have been already been
observed \citep{Wakker2008}.


Finally, we note that empirical estimates of stellar mass loss exist
for some nearby dwarf galaxies, such as LMC, WLM, and IC 1613, in
which complete samples of AGB stars can be constructed and mass loss
for individual AGB stars determined from the IR
observations \citep{jackson_etal07a,jackson_etal07b,Matsuura2009,srinivasan_etal09}.
Observational estimates of total mass loss from the AGB stars for
these galaxies range from $\sim 10\%$ (LMC) to $\sim 50-100\%$ (WLM,
IC 1613) of their total SFR. Given that additional mass loss can be
expected from supernovae and red giant stars, these independent,
direct estimates are consistent with our model calculations and also
suggest that stellar mass loss may provide substantial fuel for
ongoing star formation in these galaxies.

\section{Discussion and Conclusions}\label{sec:conclusions}

We have used a simple, empirically-motivated model for deriving star
formation histories of galaxies of different stellar masses along with mass loss rates from 
stellar evolution code to show that stellar mass loss can be the most important source of gas
for fueling continuing star formation in late type galaxies at low
redshifts ($z\lesssim 0.1-0.5$). We also explicitly show that the gas
supplied by stellar mass loss is comparable to the perceived deficit
of gas required to fuel continuing star formation in a number of
nearby galaxies. This conclusion is consistent with direct
observational estimates of stellar mass loss contribution of gas to
the ISM of nearby dwarf galaxies, such as WLM and IC1613. We presented
tests of the effects of scatter and uncertainties in the SFR-$M_\ast$
relation on our results and have show that our conclusions are not
affected.

We believe that our calculations provide a conservative estimate of
the expected stellar mass loss rates. First, they rely on star
formation histories derived from the {\it observed\/} evolution of the
SFR-$M_\ast$ relation. Thus, only if current observational results on
the evolution of this relation are significantly biased, would our
conclusions be biased as well. Moreover, in the Appendix we show that any feedback
that delays consumption of material lost by stars will only further enhance 
the rate that recycled gas gets reprocessed at late epochs.

Maintaining a gas supply for star formation at $z\approx 0$ primarily
through recycled stellar stellar mass has implications for the
recent chemical enrichment histories of galaxies. In the case of the
Milky Way, ongoing infall of pristine gas is thought to be required to
match the observed enrichment history and metallicity
distribution \citep[e.g.,][]{Sancisi2008}. However, observations of
the Milky Way and other nearby galaxies show that the metallicities of
planetary nebulae are generally lower than or equal to the
metallicities of the HII regions around young
stars \citep{stanghellini_etal10}, meaning that gas returned to the
ISM by such nebulae does not need to be diluted by infall in order to
provide fuel for currently forming stars. Furthermore, detailed
enrichment models are actually not so restrictive at late epochs. For
example, \citet{Chiappini2008} points out that although chemical
evolution models show a need for large inflow rates during the early
stages of the evolution of the Milky Way disk, a present day infall
rate of only $\approx 0.45\Msol\yri$ is used in the fiducial
\citet[][]{Chiappini2001} model that is consistent with enrichment constraints.

Deuterium abundances can potentially provide the most stringent
 constraint on the amount of gas that stellar mass loss can supply to
 galaxies like the Milky Way. Deuterium is completely destroyed in
 stars, so that the deuterium abundance in the ISM of a galaxy
 relative to its primordial value constrains the fraction of ISM
 recycled through stars \citep[see, e.g.,][for a pedagogical review
 and \citeauthor{prodanovic_etal10} \citeyear{prodanovic_etal10} for a
 recent analysis]{pagel09}. However, it is not clear how restrictive
 deuterium constraints are at present. Deuterium abundance exhibits
 variation of a factor of $\sim 3-4$ even within a kiloparsec from the
 Sun \citep{sonneborn_etal00,linsky_etal06} and longer lines of sight
 indicate lower abundances \citep{wood_etal04}. One possible
 interpretation is that variation of deuterium abundance is due to its
 depletion onto dust grains \citep{linsky_etal06}, in which case the
 largest observed abundances should be interpreted as the true
 undepleted values. Such assumption leads to conclusion that most of
 the present ISM gas in the Milky Way should be unprocessed by
 stars \citep{prodanovic_fields08,prodanovic_etal10}. However, while
 dust depletion interpretation of observed deuterium abundance
 variation is plausible, some puzzles do remain. For example, the
 correlation between deuterium abundance along a line of sight and
 reddening in the same direction, which would be expected within such
 framework, has not been detected \citep{steigman_etal07}. This
 implies that abundance variation may also be due to incomplete ISM
 mixing, in which case the constraints on recycled fraction can be
 substantially relaxed. Finally, we note also that in the context of
 this study, the deuterium abundance in the solar neighborhood may not
 be representative for the global constraints for the entire Galaxy.
 This is because most of molecular gas and star formation in the Milky
 Way occurs in a ring at $\sim 3-6$ kpc from the Galactic center and
 the deuterium abundance in that region is substantially lower than
 around the Sun \citep{Lubowich2010}.

An interesting implication of our results is that stars can be forming
largely from recycled stellar material for cosmologically significant
periods of time. The question is then why some galaxies with
significant stellar mass do not form stars during late stages of their
evolution. Although some cold gas is observed in early type
galaxies \citep[e.g.,][]{bouchard_etal05,welch_etal10}, most
elliptical galaxies have star formation rates much lower than the
expected rate of stellar mass loss. Although detailed analysis of this
issue is beyond the scope of this study, the likely explanation is
that gas lost by stars in spheroidal systems orbiting with high
velocities within tenuous hot halo is efficiently mixed with halo gas.
The situation is analogous to a well-known ``cloud in the wind''
problem \citep[e.g.,][]{agertz_etal07} with hot halo gas acting as
high-velocity wind blowing past a red giant envelope. The cloud is
expected to be unstable to the Kelvin-Helmholz (KH) instability and to
be disrupted and mixed with hot gas on a short time scale. The extent
of this mixing is a subject of active
research \citep{bregman_parriott2009, Parriott2008}. At the center of
early types, densities of recycled material should become so high that
only regular AGN activity may keep recycled material from
condensing \citep{Ciotti2007}. Interestingly, the abundance of dust
expected to come from the envelopes of red giant stars according to
stellar mass loss models is also not observed in nearby globular
clusters \citep[e.g.,][]{Barmby2009,Gnedin2002,van_loon_etal09}. In
this case, globular cluster stars are also moving through Milky Way
halo gas at high velocity, so no accumulation of cold gas shed by
stars may be possible as that gas is, again, experiencing ram
pressure, KH instabilities, and
mixing \citep[e.g.,][]{van_loon_etal09}.

The situation is likely to be different for disk galaxies because 
disk stars are releasing their gas directly into
co-rotating ISM, rather than rapidly flowing hot halo gas. The lost
gas can therefore join ISM and become readily available for star
formation. Therefore, our results and conclusions about importance of
stellar mass loss should be primarily applicable to late-type disk
galaxies. In this respect, it is interesting that observations
indicate that there is more gas in the luminous S0 galaxies than in
the elliptical galaxies of similar luminosity \citep{welch_etal10}.

Even in the disk systems, however, the stellar mass loss may only be
efficient in fueling star formation in the presence of at least some
gaseous disk. Gas lost by stars in a pure stellar disk (without any
corotating gas) would be subject to the same efficient mixing with
halo gas, as stars that are moving with high velocities through slowly
rotating halo. Therefore, stellar mass loss may become inefficient in
fueling continuing star formation when gas surface densities falls to
some small value.

The significant role of stellar mass loss in fueling continuing star
formation during late stages of evolution of galaxies implies that
it is very important to include this process in cosmological
simulations of galaxy formation. Inclusion of stellar mass loss in
simulations is straightforward: for a given choice of IMF, the mass of
gas returned to ISM in a given interval of time is given by the
differential of equation~\ref{eq:fml} with appropriate parameters (see
Table~\ref{table:fmlimf}). This mass can be added directly to a host
cell of stellar particle in a grid simulation or distributed over
neighboring gas particles in SPH simulations. Simulations with and
without stellar mass loss can be used to explore other possible
effects of this process, such as effect on galaxy
morphologies \citep{Martig2010} and their stellar masses. The latter
will depend on the star formation histories of galaxies and thus may
exhibit systematic variation with host halo mass and contribute to the
scatter of stellar mass at a fixed halo mass.

\section*{Acknowledgments}

We are grateful to Mary Putman, Oleg Gnedin, Marcel Haas, Don York and
Donald Lubowich for useful discussions and comments, to Nick Gnedin
for providing the code to generate initial conditions for the
cosmological simulations used in the appendix, and to the the anonymous
referee for helpful suggestions. This work was partially supported by
and by the NSF grant AST-0708154 and by the Kavli Institute for
Cosmological Physics at the University of Chicago through the NSF
grant PHY-0551142 and an endowment from the Kavli Foundation. The
simulations used in this work have been performed on the Joint
Fermilab - KICP Supercomputing Cluster, supported by grants from
Fermilab, Kavli Institute for Cosmological Physics, and the University
of Chicago. This work made extensive use of the NASA Astrophysics Data
System and arXiv.org preprint server. AK would like to thank the Aspen
Center for Physics and organizers of the ``Star formation in
galaxies'' workshop for hospitality and stimulating and productive
atmosphere during the completion of this paper.

\appendix
\section{Modeling Gas Reprocessing}\label{sec:reprocessing}

Until now we have dealt with mass loss from a stellar population at a particular instant, ignoring that gas returned to ISM may not be immediately accessible to form stars and may instead accumulate. Using cosmological galaxy formation simulations, we now address issue of how much gas might be available from stellar recycling to be reprocessed to form stars at low redshift, given gas reprocessing physics over galaxy's mass loss history. 

\subsection{Simulations}\label{sec:simulations}
Cosmological galaxy formation simulations of MW-sized systems
were carried out using Eulerian, gas dynamics+$N$-body adaptive refinement tree (ART)
code \citep{Kravtsov1999, Kravtsov2002, Rudd2008}. The simulations
followed the evolution of dark matter and baryons in a box of comoving
size $L_{\rm{box}}=20h^{-1}\Mpc$ in a flat $\Lambda$CDM cosmology:
$\Omega_0=1-\Omega_\Lambda=0.258,\Omega_b=0.044, h=0.72,
n_s=0.96, \sigma_8=0.80$

A standard ``zoom-in'' approach was used to generate initial
conditions in order to achieve high mass resolution, while adaptive
mesh refinement was used to increase the spatial resolution in regions
where baryons collapse to form galactic disks. Milky Way-sized systems
were selected for resimulation from a low-resolution non-radiative
simulation at $z=0$. Particles within three virial radii of the around
the center of each object were then replaced at the initial epoch
($z=62$) with eight times more high-resolution particles of mass
$m_{\rm{DM}}=2.82\times 10^7 h^{-1}\Msol$ with a spatial distribution
constrained to match the low resolution initial conditions, while also
including additional smaller scale power. There is no contamination by
low resolution dark matter particles within the virial radius of our
halos at $z=0$, but the large-scale tidal field remains accurately
sampled using this technique.

The simulation box was initially resolved by a uniform $128^3$ grid
and additional successive refinements were introduced in collapsing
regions that were occupied by the high-resolution dark matter
particles. The cells were refined if their gas mass exceeded
$m_{\rm{b}}=4.67\times 10^7h^{-1}\Msol$ or the dark matter mass in a
cell is more than $2m_{\rm{DM}}$, where $m_{\rm DM}$ is the mass of
the highest resolution dark matter particle. Cells are $610h^{-1}\pc$
across in comoving units at the maximum allowed refinement level.

Five Milky Way size halos were simulated to $z=0$. In order to isolate
 the effects of mass loss, simulations of each halo were run with mass
 loss turned on (\textit{ml}) and mass loss turned off (\textit{nml}).
 We required that the sample halos did not undergo major mergers after
 a lookback time of $5\Gyr$ and that simulations of the same halo with
 {\it ml} and {\it nml} did not differ significantly in accretion or
 timing of mergers\citep[e.g.,][]{Frenk1999} as a result of the
 differing physics. Two halos met these criteria and they are
 presented here as MW1 and MW2. They have final total masses of
 $M_{\rm{MW1}}=1.6\times 10^{12} h^{-1}\Msol$ and
 $M_{\rm{MW2}}=1.9\times 10^{12} h^{-1}\Msol$ at overdensity
 $\rho=200\rho_{\rm cr}$, where $\rho_{\rm cr}$ is the critical
 density of the universe at the epoch of analysis.

\subsubsection{Subgrid Physics}\label{sec:subgrid}

Star formation is modeled in our simulations with a simple density
dependence and efficiency:
$$\dot{\rho_*}~=~\frac{\rho_{\rm{gas}}}{\tau_\ast}~\left(\frac{\rho_{\rm{gas}}}{0.01\Msol\pc^{-3}}\right)^{0.5}$$,
and we adopt $\tau_\ast=4\Gyr$, noting that the results and our
conclusions are not very sensitive to the exact value of $\tau_\ast$.
Stellar particles are allowed to form in regions with density above
threshold value of $n_{\rm{sf}}=0.5\cmm3$ and a temperature below the
threshold value of $\rm{T}_{\rm{sf}}<9000\ \rm K$. Such a high temperature
and low density thresholds are reasonable choices for simulations of
relatively low spatial resolution that we use in this
study \citep[see][for a discussion of $\tau_\ast,n_{\rm{sf}},$ and
$\rm{T}_{\rm{sf}}$]{Saitoh2008}.

Each newly formed stellar particle is treated as an SSP with a stellar
initial mass function (IMF) that is described by the \cite{Miller1979}
functional form, with stellar masses in the range $0.1 - 100 \Msol$.
All stars with $M_{\ast} > 8 \Msol$ immediately deposit both $2 \times
10^{51}$ ergs of thermal energy, accounting for the energy input by
stellar winds and type II supernova (SN II), and a fraction $f_Z
= \min(0.2, \frac{0.01M_{\ast}}{\Msol} - 0.06)$ of their mass as
metals, into the surrounding gas, in order to approximate the results
of \cite{Woosley1995}. The code also accounts for the SN Ia feedback,
assuming a rate that slowly increases with time and broadly peaks at
the population age of $1 \Gyr$. We assume that a fraction of
$1.5 \times 10^{-2}$ of mass in stars between 3 and 8 $\Msol$ explodes
as SN Ia over the entire population history and that each SN Ia dumps
$2 \times 10^{51} \erg$ of thermal energy and ejects $1.3 \Msol$ of
metals into the surrounding gas. For the assumed IMF, 75 SN II
(instantly) and 11 SN Ia (over several billion years) are produced by
a $10^4 \Msol$ stellar particle. This sort of model, without ad hoc
kicks or delayed cooling, has been shown to have a negligible effect
on the SFH and resulting galaxy properties compared to a model without
feedback \citep{Katz1992,tassis_etal08,Schaye2010}. Metallicity
dependent equilibrium cooling rates, as well as UV heating rates, due
to the cosmological ionizing background, are tabulated using the
Cloudy code \citep[][v96b4]{Ferland1998}.

The stellar mass loss of each stellar particle is modeled using eq.~\ref{eq:fml}
with $C_0=0.05$ and $\lambda=5\Myr$ and  
about $40\%$ of initial mass of
 stellar particle is lost over a Hubble time with these parameters.
 At each time step, gas mass lost by a stellar particle is added to
 gas of its parent cell along with its energy and momentum. The
 interpretation of our simulation results is not sensitive to the
 specific choice of $C_0$ and $\lambda$.

As a result of the known problems with simple galaxy formation
 simulations, our galaxies overcool and produce stellar fractions that
 are about $5$ times too high compared to observational expectations.
 To account for this, we re-normalize all star formation histories of
 the simulated galaxies with the constant ratio of the final stellar mass in
 the {\it ml} simulation to the Milky Way value of
 $5.5\times10^{10}\Msol$ \citep[e.g.,][]{Flynn2006}.

\subsection{Gas Reprocessing Results}\label{sec:reprocessingresults}

\begin{figure*}[!t]
\begin{center}
\resizebox{7in}{!}{ \includegraphics{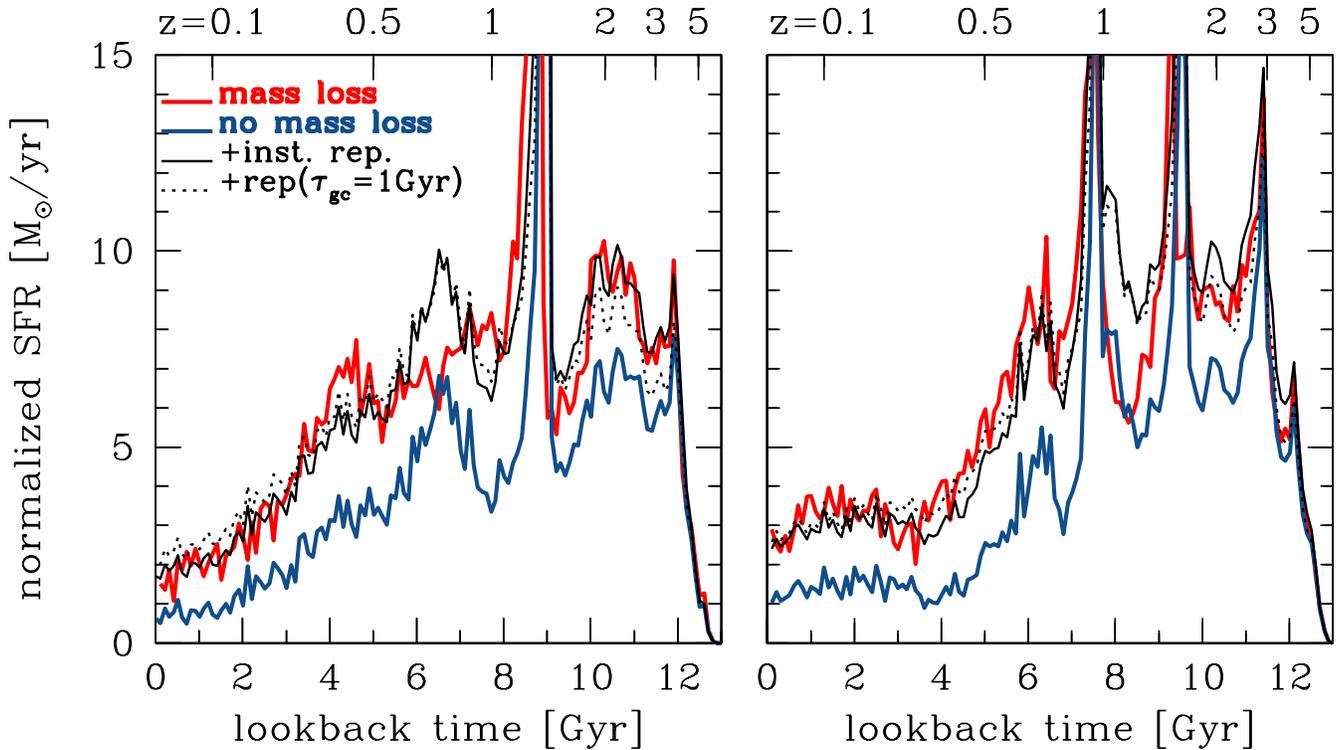} }
\end{center}
\caption{
Simulated SFHs that include all stars within $30h^{-1}\Kpc$ of the
 galaxy and are normalized in such a way that the final stellar mass
 is equal to $5.5\times10^{10}\Msol$. The thin blue lines are SFHs for
 simulations that do not include mass loss from stars (\textit{nml}),
 while the thick red line is a fiducial simulation with mass loss
 included (\textit{ml}). Black lines are gas recycling models, with
 gas consumption timescales $\tgc = 0 \Gyr$ (solid) and $\tgc =
 1 \Gyr$ (dotted), that attempt to map the \nml SFH onto the \ml SFH.
 }
\label{fig:sfhml} 

\end{figure*}

The star formation histories of our two simulated galaxies are plotted
in Figure~\ref{fig:sfhml}. There are no major mergers after $z=0.5$,
and these SFHs show few sustained bursts, making their growth largely
consistent with expectations for the build up of the Milky
Way \citep[e.g.,][]{Rocha-Pinto2000}. The SFHs show that $z=0$ SFRs in
the runs with mass loss (\ml plotted in red) are more than a factor of
two higher than the SFRs in the corresponding simulations without mass
loss (\nml plotted in blue). This confirms the basic expectation that
gas lost by stars can be a significant source of fuel for star
formation at low redshift.

We use the \nml simulation to test whether simple assumptions about
 the fate of gas lost by stars can reproduce results of simulations
 with stellar mass loss. The GMLR is given by eq.~\ref{eq:grr}, however,
the gas returned to the ISM can accumulate and be consumed over a
 certain gas consumption time scale. The mass that accumulates in the
 ISM from stellar mass loss between time $t_0$ and $t$ is
 then \begin{equation} \label{mmlt} m_{\rm{ml}}(t_0,t)
 = \int_{t_0}^t\left\{\GMLR(t')- \left[ \SFR(t')-
\SFR_{\rm nml }(t') \right]\right\} dt'
 \end{equation} To determine the SFR and hence a stellar mass loss
rate, we assume a linear SFR scaling with gas density that has a gas
consumption time scale $\tgc$. Enhancement in star formation is
therefore directly proportional to mass added to the star forming ISM,
\begin{equation}
\SFR(t)=\SFR_{\rm nml}(t)+\frac{m_{\rm{ml}}(0,t)}{\tgc}  
\label{eq:sfrlin}
\end{equation}
Stars formed from the gas lost by stars will themselves lose mass, and
this mass will, in turn form stars. In order to fully account for the
additional star formation resulting from mass loss,
equation~\ref{eq:sfrlin} needs to be iterated when starting from
the \nml case, replacing $\SFR_{\rm nml}(t)$ with each subsequent
$\SFR(t)$, until $\SFR(t)$ converges.

The simulation's true star formation law is nonlinear in density,
 meaning that $\tgc$ effectively varies with density. The average gas
 densities of a star forming disk in simulations can vary from
 $\tgc=500\Myr$ to a $\tgc$ of few billion years, with stars tending
 to form in lower density gas (longer $\tgc$) as accretion slows and
 the gas in their disks is depleted. However, this change in the
 efficiency of gas conversion occurs slowly and, furthermore, the
 effect of $\tgc$ is mitigated by the fact that a reduction in $\tgc$
 results in a build up of $m_{\rm{ml}}(0,t)$, which leads to more star
 formation. A constant $\tgc$ therefore mimics the relevant behavior.

An alternative, simpler model, which we will call {\it instant reprocessing}
\footnote{Instant reprocessing is distinct from the instant recycling approximation. Instant {\it recycling} approximates mass loss as instantaneous \citep[e.g.,][]{Tinsley1980}, whereas instant {\it reprocessing} approximates the birth of stars from returned material as instantaneous.}, 
 can be constructed by taking $\tgc$ to $0$. Then gas returned to ISM
 by stars is immediately consumed by star formation so only ongoing
 mass loss contributes to star formation and thus
\begin{equation}
\SFR(t)=\SFR_{\rm nml}(t)+\GMLR(t).
\end{equation}

We plot SFHs with both a $\tgc = 1 \Gyr$, and instantaneous
reprocessing applied to the SFH from the \nml simulation in
Figure~\ref{fig:sfhml}. The mapping between simulations is not exact.
Gas returned during a merger may be returned to the diffuse halo and
be temporarily lost to the star forming disk. Nonetheless, throughout
most of its lifetime, and especially in late epochs when the galaxy is
quiescently evolving, the simple assumption of instant reprocessing
applied to the \nml simulations generates an excellent match to the
SFH in the \ml simulations. 

Reprocessing models with longer gas consumption timescales result in gas accumulation and
more gas reprocessing at late times, because they smooth the declining
reprocessing rate curve. For the simulated SFHs, the effect of the
$\tgc = 1 \Gyr$ reprocessing is a $\lesssim10\%$ enhancement in SFRs
at late epochs compared to the instantaneous reprocessing model. It is
worth emphasizing that any feedback mechanism that slows the
processing of gas to stars would have the effect of raising $\tgc$ and
would therefore imply a higher rate of gas reprocessing at late
epochs. 

\bibliography{lit.bib}

\end{document}